\documentclass[12pt]{article}
\usepackage{arxiv}


\normalsize

\usepackage{amsmath}
\usepackage{graphicx}
\usepackage{booktabs,tabularx,array,multirow}
\usepackage{microtype}
\usepackage{caption}
\usepackage{pdfpages}
\usepackage[hidelinks]{hyperref}
\usepackage[style=apa,backend=biber,sorting=nyt]{biblatex}
\let\cite\parencite

\renewcommand{\shorttitle}{SOC indicators and large language models}
\renewcommand{\headeright}{Cheng et al.}

\title{\bfseries How Reliable Are Large Language Models for Soil Organic Carbon Assessment? An Australian Benchmark}

\author{
Yingxuan Cheng$^{1}$, Tong Li$^{2,*}$, Yayong Li$^{1}$, Yuchuan Fan$^{3,4}$, You-Gan Wang$^{1}$ \\[4pt]
\textbf{Jinran Wu$^{1,*}$, Scott Chapman$^{2}$, Geoffrey J. McLachlan$^{1}$}\\[8pt]
\normalsize $^{1}$School of Mathematics and Physics, The University of Queensland, QLD 4072, Australia\\
\normalsize $^{2}$School of Agriculture and Food Sustainability, The University of Queensland, QLD 4072, Australia\\
\normalsize $^{3}$College of Jiyang, Zhejiang A\&F University, Zhuji 311800, China\\
\normalsize $^{4}$Department of Soil, Water, and Ecosystem Sciences, University of Florida-IFAS, FL 32611, USA\\[4pt]
\normalsize $^*$Correspondence: tong.li1@uq.edu.au; jinran.wu@uq.edu.au
}

\date{}

\begin{document}
\maketitle

\begin{abstract}
Soil organic carbon (SOC) is an indicator of agricultural soil condition and carbon storage, but numerical outputs from large language models (LLMs) require validation before use in monitoring or management. We evaluated seven general-purpose LLM configurations on three direct-response tasks: estimating 0--10 cm SOC concentrations for 308 Australian paddocks under two information conditions, calculating SOC stock in 120 matched scenarios, and comparing 300 paddock pairs. References were measured SOC records or deterministic calculations; response validity, repeated-call dispersion and estimated API cost were assessed separately. Additional measured soil attributes reduced concentration mean absolute error by 13--62\% across configurations, although high-SOC underprediction and paddock-specific errors persisted. Six configurations were highly accurate in stock calculations using equation-native units without coarse fragments, but accuracy declined with physically equivalent alternative units. Pairwise accuracy was 90.5--94.6\% overall and 59.1--70.9\% for 52 closely separated pairs; some configurations were sensitive to presentation order. Nearly all outputs were parseable, without necessarily being correct. The benchmark identifies three distinct requirements for AI-mediated SOC assessment: agreement with measured concentrations, unit-consistent stock calculations and reliable discrimination between sites. These findings support task-specific validation and auditable calculations, rather than unverified use of LLM outputs as ecological evidence.
\end{abstract}


\section{Introduction}

Soil organic carbon (SOC) concentration and stock are widely used in agricultural monitoring, carbon accounting and land management \cite{main_ref_030,main_ref_034,main_ref_038}. Concentration describes carbon per unit dry soil mass; stock expresses carbon per unit area within a defined layer. Although SOC is related to nutrient cycling and soil structure, it is not a complete measure of soil health, which also depends on physical and biological functions \cite{main_ref_010,main_ref_022,main_ref_056}. Interpreting either SOC quantity requires attention to soil context, sampling depth and intended use \cite{main_ref_011,main_ref_019}. Errors in concentration estimates or stock conversion can therefore distort a monitoring result without any underlying ecological change.

Interpretation also depends on reference systems. SOC benchmarking
methods can yield different classifications for the same soil, and
regional variation reflects pedogenic conditions as well as management
\cite{main_ref_003,main_ref_015}. Consequently,
acceptable error and useful discrimination depend on whether a value
supports exploratory screening, local management or formal stock
reporting. Evaluating an AI-generated SOC value requires more than
numerical plausibility, but it must not imply that SOC alone establishes
ecosystem services or management impacts.

General-purpose large language models (LLMs) can produce numerical
answers from textual instructions without task-specific fitting \cite{main_ref_005}. Their use in soil science and agricultural advice is
expanding \cite{main_ref_006,main_ref_007,main_ref_009}, yet fluent responses may be inaccurate \cite{main_ref_023}. Unlike
established SOC prediction approaches evaluated against field
observations, such as spectroscopy and spatial modelling \cite{main_ref_031,main_ref_032,main_ref_048}, LLM applications have primarily been
assessed for workflow assistance, literature synthesis or hypothesis
generation \cite{main_ref_009,main_ref_028,main_ref_036}. The validation scope of conventional soil modelling also remains
important when interpreting apparent predictive skill \cite{main_ref_008,main_ref_014}.

Existing LLM benchmarks cover scientific calculation, numerical
reasoning and agricultural knowledge \cite{main_ref_025,main_ref_029,main_ref_052,main_ref_054,main_ref_055}, including
soil-science questions and ecological knowledge \cite{main_ref_013,main_ref_026}. Their findings cannot be transferred directly to
field-referenced SOC estimates. Results on regression and geographic
tasks depend on the examples, contextual inputs and spatial information
provided \cite{main_ref_033,main_ref_040,main_ref_049,main_ref_050}. A SOC-specific evaluation must instead
specify the reference quantity and test the conditions under which an
output remains correct.

Three potential failure modes are particularly relevant to SOC
monitoring: disagreement with measured concentration, inconsistent stock
calculations under equivalent unit representations, and unreliable
ordering of paddocks with similar SOC. These operations have different
references and error consequences, so a single aggregate score would
obscure their limitations. Broader evidence that soil-health assessments
can disagree even for identical sites reinforces the need to evaluate
indicator performance against its intended purpose \cite{main_ref_058}.

We benchmarked seven LLM configurations on concentration estimation
(T1), stock calculation (T2) and pairwise comparison (T3). T1 evaluated
direct estimates against measured 0--10 cm references and assessed the
effect of additional soil measurements; T2 tested execution of a
supplied stock equation under matched unit and coarse-fragment
conditions; T3 assessed ordering accuracy by SOC separation and
presentation position. Independent API calls excluded retrieval,
external calculators and conversation history \cite{main_ref_043,main_ref_044}. T1 and T3 drew on 308 Australian paddocks in
SOC-M \cite{main_ref_024,main_ref_039}, whereas T2 used
deterministic synthetic cases. Spatially validated supervised baselines
contextualised T1, without being equivalent closed-book comparators. The analysis asks when additional data improve estimation, whether
physically equivalent unit expressions preserve calculation accuracy,
how separation and order affect comparison, and whether validity, output
variability and cost add independent operational information. It
provides a baseline for testing future retrieval- or calculator-assisted
workflows, not a replacement for SOC measurement or a composite model
ranking.

\section{Methods}

\subsection{Data and reference SOC values}

\begin{figure}[!htbp]
\centering
\includegraphics[width=0.95\linewidth]{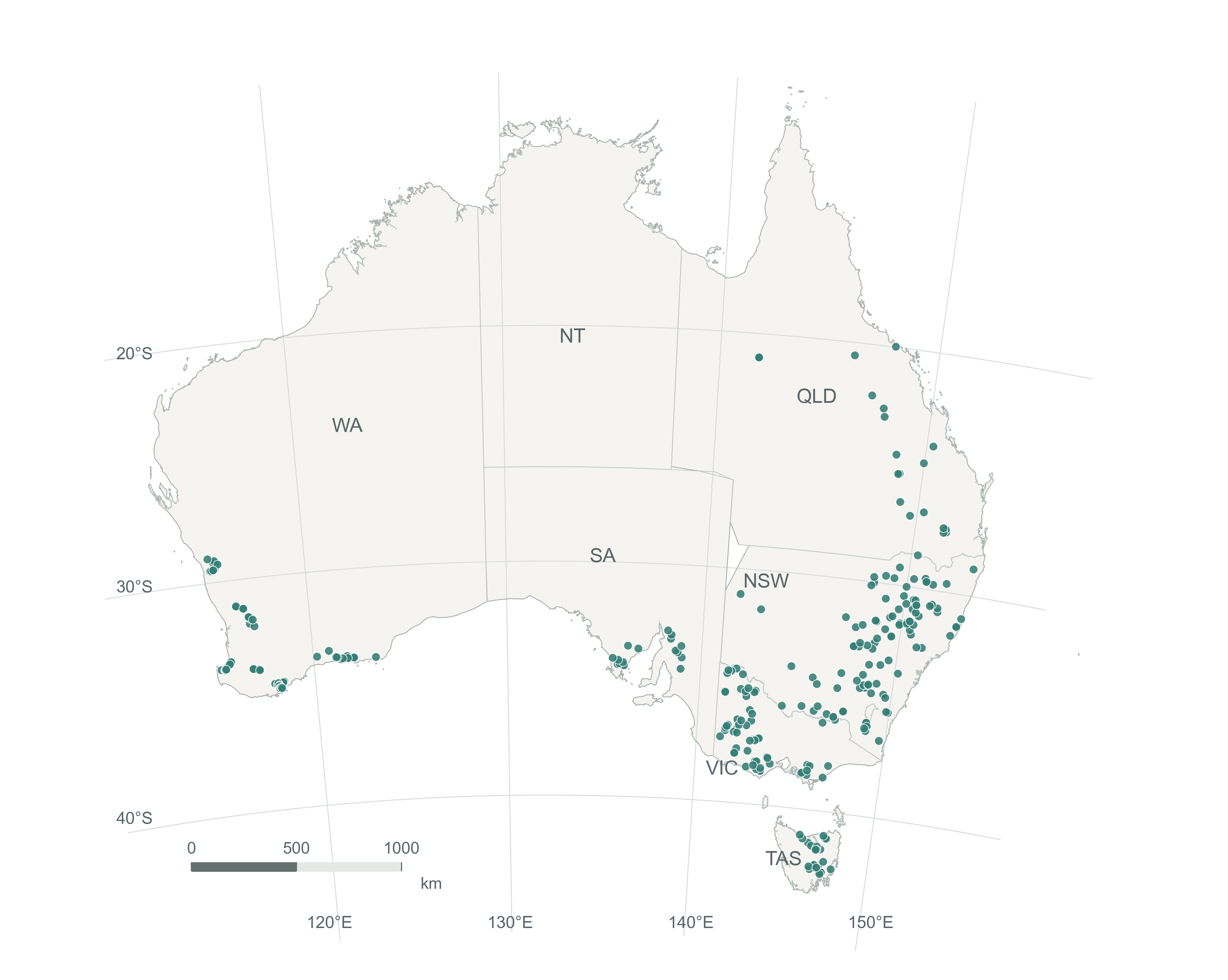}
\caption{Geographical distribution of the study paddocks in
Australia. Points show the dataset-reported sampling locations of the
308 paddocks with 0--10 cm reference SOC values across six Australian
states; they do not mark paddock centroids or boundaries, and colour
does not encode SOC concentration. No sampling point was moved or
omitted. Coordinates were projected to Australian Albers (EPSG:3577),
and coastline and state polygons were simplified for display. Data
sources: SOC-M v1 \cite{main_ref_024}, CSIRO Data Access Portal,
DOI: 10.25919/z378-4286, © CSIRO 2025; and Australian Bureau of
Statistics (ABS) Australian Statistical Geography Standard (ASGS)
Edition 3 States and Territories 2021, © Commonwealth of Australia. Both
datasets are used under CC BY 4.0
(\url{https://creativecommons.org/licenses/by/4.0/}).}
\label{fig:main1}
\end{figure}

The empirical data were obtained from version 1 of the Soil Organic
Carbon Monitoring (SOC-M) dataset \cite{main_ref_024,main_ref_039}. SOC-M resampled 308 agricultural sites across six Australian
states during 2022--2025, revisiting locations surveyed in 2009--2012
(Fig.~\ref{fig:main1}). The sampling design targeted ten soil cores per site,
generally within a 25  $\times$  25 m grid, with larger sampling areas at some
Queensland rangeland sites. Samples were analysed individually or
composited by depth according to the sampling design \cite{main_ref_039}.

The benchmark retained 1,213 analytical records from the 0--10 cm layer
across all 308 sites. These sites are hereafter referred to as paddocks,
although each reference represents the sampled area rather than the
entire field. Each reference SOC concentration was the unweighted
arithmetic mean of the paddock's retained records, expressed in g C
kg\textsuperscript{-1} on an oven-dry soil mass basis. Paddock counts
retained record counts and reference SOC distributions are summarised by
state in Table~\ref{tab:main1}. Reference concentrations ranged from 1.1 to 93.0 g C
kg\textsuperscript{-1}, with a median of 18.0 g C
kg\textsuperscript{-1}. State-specific medians ranged from 11.8 g C
kg\textsuperscript{-1} in Queensland to 39.7 g C kg\textsuperscript{-1}
in Tasmania. These descriptive differences provide context for the
prespecified same-state median baseline. Retained records per paddock are summarised in Table~\ref{tab:main1}.

\begin{table}[htbp]
\centering
\caption{Paddock-level reference SOC concentrations and retained
analytical records for the 0--10 cm layer by Australian state}
\label{tab:main1}

\begin{tabular}{lrrrrrrrr}
\toprule
State & Paddocks & Records & Records/site & Mean & SD
& Median & Q1--Q3 & Min--max \\
\midrule
NSW & 103 & 363 & 3.5 & 20.1 & 10.2 & 18.3 & 14.5--24.0 & 3.0--69.1 \\
WA  & 79  & 326 & 4.1 & 23.0 & 20.0 & 14.1 & 5.9--35.7  & 3.0--72.4 \\
VIC & 64  & 210 & 3.3 & 30.4 & 23.1 & 22.1 & 13.5--44.7 & 1.1--93.0 \\
TAS & 25  & 160 & 6.4 & 42.1 & 16.9 & 39.7 & 29.0--51.8 & 15.7--82.4 \\
QLD & 21  & 93  & 4.4 & 11.6 & 4.3  & 11.8 & 9.7--14.1  & 4.8--18.7 \\
SA  & 16  & 61  & 3.8 & 15.2 & 5.2  & 13.6 & 12.4--16.6 & 10.0--30.2 \\
\midrule
All & 308 & 1,213 & 3.9 & 24.0 & 18.0 & 18.0 & 11.8--31.3 & 1.1--93.0 \\
\bottomrule
\end{tabular}

\par\smallskip
\begin{minipage}{\linewidth}
\footnotesize
\textit{Note.} SOC concentrations are expressed in
g C kg$^{-1}$. Each paddock contributes one reference value,
calculated as the unweighted arithmetic mean of its retained
0--10 cm analytical records. Records/site is the retained
record count divided by the paddock count. SD denotes the
sample standard deviation; Q1--Q3 gives the 25th and 75th
percentiles. Overall SOC statistics were calculated directly
across all 308 paddocks, rather than by averaging state-level
summaries. NSW, New South Wales; WA, Western Australia;
VIC, Victoria; TAS, Tasmania; QLD, Queensland;
SA, South Australia.
\end{minipage}

\end{table}

Dataset-level laboratory documentation describes high-temperature
oxidative combustion of finely ground soil from the $<$2 mm
fraction, with inorganic carbon removed from carbonate-bearing samples
where required \cite{main_ref_017,main_ref_039}. The released
analytical table lacked a record-specific analytical-method field,
preventing independent verification for every retained record. No
inter-method conversion was applied.

\subsection{Model configurations and evaluation protocol}

\begin{figure}[!ht]
\centering
\includegraphics[width=\linewidth,height=0.64\textheight,keepaspectratio]{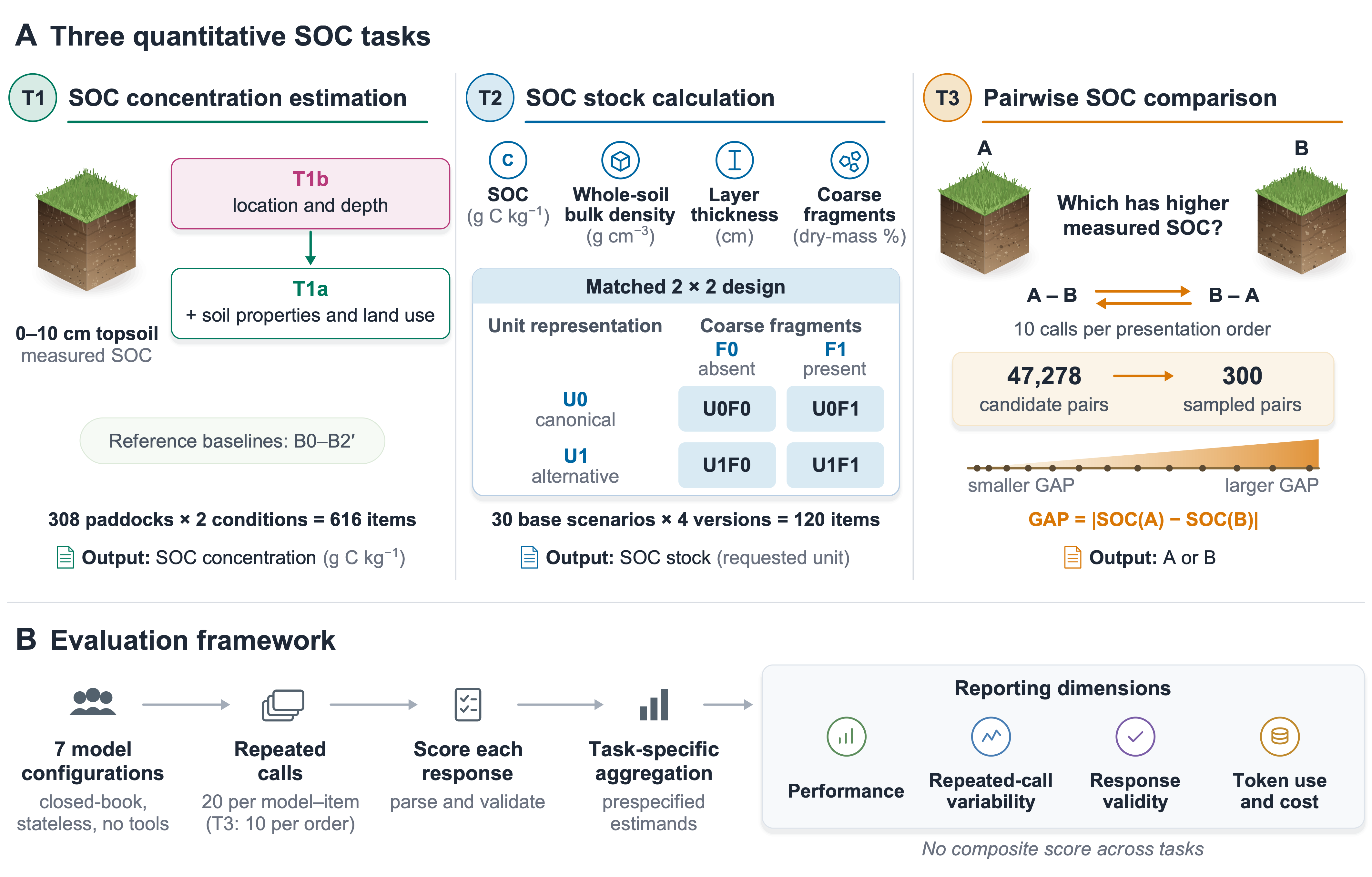}
\caption{Benchmark design and evaluation framework. (A) T1
evaluated SOC concentration estimation for 308 paddocks under two nested
information conditions: T1a added soil properties and historical land
use to the location and depth information in T1b; B0, B1, B2 and B2$^{\prime}$ are
prespecified supervised reference baselines. T2 crossed a
single-dimension unit-representation condition (U0/U1) with a
coarse-fragment dry-mass condition (F0/F1) across 30 base scenarios,
yielding 120 items; each response gave SOC stock in the requested unit.
T3 sampled 300 pairs without replacement from 47,278 eligible unordered
pairs and presented each pair ten times in each presentation order; GAP,
the absolute difference in reference SOC between the two paddocks, was
analysed as a continuous variable. (B) Seven configurations were
evaluated through stateless, closed-book calls without tools or
retrieval. Responses were scored before task-specific aggregation.
Performance, repeated-call variability, response validity, token use and
estimated API cost were reported separately, without a cross-task
composite score. The total of 145,040 refers to planned responses.}
\label{fig:main2}
\end{figure}

Seven configurations from six providers were evaluated (Table~\ref{tab:main2}). A
configuration denotes the requested model ID and recorded settings;
provider-side versions could not be independently verified. The
benchmark comprised 616 T1 items, 120 T2 items and 300 T3 pairs (Fig.~\ref{fig:main2}). Each configuration had 20 planned responses per T1 or T2 item and 20
per T3 pair, split equally between presentation orders, giving 145,040
planned responses.

\begin{table}[htbp]
\centering
\caption{Evaluated model configurations}
\label{tab:main2}
\begin{tabular}{lllcc}
\toprule
Configuration & Provider & Model ID & Reasoning & Token limit \\
\midrule
GPT-5.6 Sol        & OpenAI    & gpt-5.6-sol       & none     & 1,024 \\
GPT-5.6 Terra      & OpenAI    & gpt-5.6-terra     & none     & 1,024 \\
Claude Sonnet 5    & Anthropic & claude-sonnet-5   & disabled & 1,024 \\
Gemini 3.8 Flash   & Google    & gemini-3.8-flash  & low      & 1,024 \\
Grok 4.6           & xAI       & grok-4.6         & low      & 1,024 \\
DeepSeek V4 Pro    & DeepSeek  & deepseek-v4-pro   & disabled & 1,024 \\
Mistral Medium 3.5 & Mistral   & mistral-medium-3-5 & none    & 1,024 \\
\bottomrule
\end{tabular}
\par\smallskip
\begin{minipage}{\linewidth}
\footnotesize
\textit{Note.} Settings are those recorded in the
request bodies. Token-limit definitions differed among provider
interfaces, so identical numerical limits do not imply
identical visible-output budgets.
\end{minipage}
\end{table}

Each response was requested through a separate application programming
interface (API) call without conversation history, retrieval or tool
access. T1 and T3 prompts included coordinates but omitted dataset
names, internal paddock identifiers and reference SOC values.
Temperature, top-p and seed were not specified. Reasoning controls were
explicitly set to none or disabled for five configurations and low for
Gemini 3.8 Flash and Grok 4.6. At the time of collection, low was the
lowest setting supported by these two interfaces. These settings do not
establish equivalent internal computation across providers. All requests
specified a 1,024-token generation limit, subject to provider-specific
counting conventions; Gemini's limit included both thought and candidate
tokens.

Responses were collected between 29 September and 5 October 2026 using a
fixed interleaved, serial schedule. Operational failures could trigger
retries of the same request, but answers were not retried because they
were incorrect or unparseable. Exact prompts, request fields,
token-limit definitions and collection details are described in the Methods section.

\subsection{Quantitative SOC tasks}

The tasks assessed three complementary operations: agreement of
estimated concentration with measured SOC (T1), correct stock
calculation across unit and fragment conditions (T2), and accurate
paddock comparison across SOC separations and presentation orders (T3).
Their distinct references, metrics and statistical contrasts are
summarised in Table~\ref{tab:main3}; task scores were not combined.

\begin{table}[!ht]
\centering
\caption{Benchmark tasks, performance metrics and statistical contrasts}
\label{tab:main3}

\begin{tabularx}{\textwidth}{lXXXX}
\toprule
Task & Benchmark items & Primary metric & Main contrasts & Uncertainty \\
\midrule

T1 &
308 paddocks $\times$ 2 conditions (616 items);
20 responses per configuration--item;
86,240 planned responses &
MAE (g C kg$^{-1}$): absolute errors averaged within
paddock and condition, then equally across paddocks &
Paired T1b $-$ T1a error difference;
supervised baselines B0--B2$'$ &
Point estimates only \\

\midrule

T2 &
30 base scenarios $\times$ 4 matched versions (120 items);
16,800 planned responses &
Accuracy (absolute relative error $\leq 2\%$):
mean binary score across 20 responses per item,
then equally across items &
Matched U1 $-$ U0; F1 $-$ F0;
U $\times$ F interaction &
Pointwise repeated-call 95\% intervals \\

\midrule

T3 &
300 paddock pairs $\times$ 2 presentation orders;
42,000 planned responses &
Direct accuracy: mean binary score within each order,
then equal weighting of orders and pairs &
21 pairwise configuration differences;
GAP and position associations &
Pointwise repeated-call 95\% intervals;
pointwise Wald intervals for secondary estimates \\

\bottomrule
\end{tabularx}

\par\smallskip
\begin{minipage}{\textwidth}
\footnotesize
\textit{Note.} Response counts refer to planned responses,
not request attempts or valid outputs (total: 145,040).
Repeated-call intervals are conditional on the fixed
benchmark items and do not include uncertainty from
sampling additional paddocks, scenarios or pairs.
\end{minipage}

\end{table}

\subsubsection{T1: SOC concentration estimation}

T1 requested one SOC concentration in g C kg\textsuperscript{-1} for
each paddock's 0--10 cm layer under two nested information conditions.
T1b supplied state, coordinates, depth interval and, where available,
the Grains Research and Development Corporation (GRDC) region. T1a
retained these fields and added historical 2012 land use, whole-soil
bulk density, coarse-fragment dry-mass fraction, total nitrogen,
electrical conductivity and CaCl$_2$ pH. Additional information was evaluated as one block, rather than through individual-field ablation. Missing fields were omitted without imputation; the benchmark used fixed prompt templates.

Absolute errors of valid finite predictions were averaged within each
configuration--paddock--condition cell and then equally across paddocks
to obtain mean absolute error (MAE). The information contrast was the
within-paddock difference in mean error, T1b - T1a, averaged equally
across paddocks; positive values indicate lower error under T1a.
Relative MAE reductions were expressed as 100  $\times$  (MAE\_T1b -
MAE\_T1a)/MAE\_T1b.

Four supervised baselines contextualised T1: the training-fold SOC
median (B0); the same-state training-fold median, falling back to B0
where needed (B1); and random forests using T1a and T1b predictors (B2 and B2$^{\prime}$, respectively; \cite{main_ref_004}). All used five-fold spatial
cross-validation with coordinate-based, SOC-blind clusters, and
preprocessing was fitted only on training folds. Withheld geographical
groups make these spatially held-out predictions rather than
random-split results \cite{main_ref_042,main_ref_051}. The
supervised baselines used measured training labels and were not
like-for-like closed-book comparators (Supplementary Methods M3; Table S6).

Figure~\ref{fig:main3} reports paddock-level mean errors and paired contrasts. Figure~\ref{fig:main4} shows the mean valid prediction for each paddock--condition cell
against its reference, with descriptive linear fits. T1 MAE was
calculated from individual responses, not from averaged predictions; the
plots were added after data collection.

\subsubsection{T2: SOC stock calculation}

T2 assessed execution of a supplied equation using 30 synthetic base
scenarios rather than empirical SOC-M records:
\begin{equation}\label{eq:stock}
\text{S}=\text{C} \times \text{BD} \times \text{d} \times (1-\text{CF})\times 0.1
\end{equation}
where S is SOC stock (t C ha-1), C is SOC concentration (g C kg-1), BD
is whole-soil bulk density (g cm-3), d is layer thickness (cm), and CF
is the coarse-fragment dry-mass fraction. Coarse fragments were
displayed as a percentage and converted to a fraction for calculation.
Under these definitions, BD(1 - CF) represents fine-soil dry mass per
unit bulk-soil volume \cite{main_ref_041}. The physical
interpretation of Eq.~\eqref{eq:stock} as fine-soil SOC stock requires C to be
expressed per unit dry fine-soil mass. T2 evaluated execution of the
supplied calculation convention, with CF defined as a dry-mass fraction
rather than a volumetric fraction.

Each base scenario had four matched versions: U0F0, U1F0, U0F1 and U1F1.
U0 used the equation's units; U1 re-expressed one input or the output
unit without changing the physical stock at a fixed fragment level. F0
specified no coarse fragments, whereas F1 supplied a non-zero mass
fraction and changed the reference stock. All 120 references used the
displayed, rounded prompt quantities. Thus, U tested representational
invariance and F tested application of the specified fragment
correction.

A valid finite response was scored as correct if its unrounded absolute
relative error from the deterministic reference did not exceed the
prespecified tolerance of 2\%; invalid or unavailable responses were
scored as incorrect. For each configuration, accuracy was calculated by
averaging binary scores across the 20 planned responses per item and
then equally across all 120 items. Unit-representation (U1 - U0) and
coarse-fragment (F1 - F0) contrasts were calculated within each base
scenario by averaging matched differences over the other factor. The U $\times$ F
interaction was defined as
\begin{equation}\label{eq:interaction}
 (p_{11}-p_{01})-(p_{10}-p_{00})
\end{equation}
where \(p_{uf}\) denotes item accuracy under \(U = u\) and \(F = f\)
within a base scenario. All contrasts were averaged equally across the
30 base scenarios and reported in percentage points. Unit-representation
and coarse-fragment contrasts are shown in Figure~\ref{fig:main5}B and \ref{fig:main5}C,
respectively, and interaction contrasts are reported in the Supplementary Methods (M4; Table S4).
Scenario ranges, unit variants and prompts are described in the Methods section.

\subsubsection{T3: Pairwise SOC comparison}

T3 tested identification of the paddock with the higher reference SOC. A
simple random sample of 300 pairs was drawn without replacement from
47,278 eligible unordered pairs, without GAP stratification. Each
paddock was described using its available T1a fields. The higher-SOC
paddock appeared as A in ten planned responses and B in the other ten.
GAP was the absolute difference between reference SOC concentrations in
g C kg\textsuperscript{-1}. Direct accuracy averaged binary correctness
within each order, weighted the two orders equally within pairs, and
then weighted pairs equally.

A secondary grouped-binomial model with a logit link related correctness
to continuous GAP and correct-answer position. A three-degree-of-freedom
natural cubic spline represented log(1 + GAP), with
configuration-specific spline and position terms and penalised terms
accounting for shared-paddock, pair and configuration-by-pair
dependence. No GAP-by-position interaction was included on the logit
scale. A separate prespecified sensitivity analysis reported accuracy
for 248 pairs meeting a nominal separation threshold and the remaining
52 pairs. The threshold used a nominal 10\% relative-error convention
for each reference; it neither quantified measured reference uncertainty
nor determined primary pair selection. Direct accuracy and conditional
associations are displayed separately in Fig.~\ref{fig:main6}. Full specifications are
provided in the Supplementary Methods (M5).

\subsection{Response validity, repeated-call variability and estimated cost}

Outputs were scored for one task-specific answer: a finite number for T1
and T2, or A/B for T3. Strict JSON parsing and prespecified fallback
acceptance were recorded separately, without reference-assisted repair.
Invalid content was distinguished from terminally unavailable responses.
Only valid finite predictions contributed to T1 errors; invalid and
unavailable responses counted as incorrect in T2 and T3. Validity was
calculated over all planned responses (Supplementary Methods M6).

Additional descriptive summaries of repeated outputs were constructed
after data collection (Fig.~\ref{fig:main7}A). T1 dispersion was the interquartile
range (IQR) of valid predictions within each
configuration--paddock--condition cell. T2 dispersion was the IQR of
valid finite numerical answers within each configuration--item cell,
expressed as a percentage of the absolute reference stock:
\begin{equation}\label{eq:dispersion}
 D_{\mathrm{T2}}=100\times\frac{Q_{(0.75)}(\hat S)-Q_{(0.25)}(\hat S)}{|S_{\mathrm{ref}}|}.
\end{equation}
Predictions and references used matching units. Valid but incorrect
answers were retained. T3 disagreement was calculated separately within
each configuration--pair--presentation-order cell:
\begin{equation}\label{eq:disagreement}
D_{\mathrm{T3}}=100\times\left[1-\frac{\max(n_{\mathrm A},n_{\mathrm B})}{n_{\mathrm{valid}}}\right]
\end{equation}
where \(n_{A}\) and \(n_{B}\) count valid A and B choices, respectively,
and $n_{\mathrm{valid}}=n_A+n_B$. Orders were not pooled when calculating
disagreement. Cell-level distributions were displayed with equal cell
weights; cells with fewer than two valid responses were unavailable, not
assigned zero. Quartiles were calculated before display transformations.
Zero numerical IQR does not establish identical answers, and zero choice
disagreement does not establish correctness. These task-specific
measures were not combined into a common stability score. Further
implementation details are provided in Supplementary Methods M7.

API cost was estimated from interpretable provider-reported token usage
across all attempts, including retries, using the fixed input and output
rates recorded on 29 September 2026. Separate cached-input discounts
were not applied. Reasoning tokens already included in output totals
were not counted again. Missing or uninterpretable usage retained a
separate unknown-cost reservation rather than being priced at zero.
Reported costs were standardised rate-based estimates, not reconstructed
invoices, and
the recorded input and output rates are listed in Table S14.

\subsection{Statistical analysis}

Inference concerned performance on fixed benchmark items, separating
finite repeated-response uncertainty from variation across potential new
sites or scenarios \cite{main_ref_035}. T2 and T3 intervals quantify
repeated-call uncertainty conditional on those items and reference
values, not performance on new monitoring sites. T1 was reported
descriptively, without confidence intervals or tests.

Bootstrap resampling used 20 planned binary outcomes within each T2 item
and ten within each T3 pair--presentation-order cell, retaining all
items, conditions and aggregation weights across 10,000 replicates
\cite{main_ref_016}. Invalid and unavailable outcomes remained
zero scores. Repeated responses were treated as exchangeable within
cells, without pairing repeat indices across cells. Bootstrap deviations
from the observed estimate were multiplied by
\(\sqrt{n\text{/}(n - 1)}\), with \emph{n} = 20 for T2 and \emph{n} = 10
for T3, to account for finite-sample variance shrinkage under empirical
resampling \cite{main_ref_020}. Pointwise 95\% intervals used the
2.5th and 97.5th percentiles of the rescaled distributions
(Supplementary Methods M6).

The 21 prespecified T3 between-configuration accuracy differences used
$Z = \Delta / \mathrm{SE}(\Delta)$, with each configuration's variance
estimated as
$\sum [p(1-p)/9]/600^2$
over its 600 pair--position cells ($p$ = observed ten-call cell accuracy)
and the contrast SE as the square root of the two configuration variances
summed, without a cross-configuration covariance term.
Two-sided normal-approximation $P$ values were adjusted as one family
by Holm's step-down procedure \cite{main_ref_021}.
This adjustment did not alter individual pointwise intervals.
The secondary model in Section~2.3.3 provided conditional GAP predictions
and position contrasts with random effects set to zero.
Pointwise 95\% Wald intervals were conditional on estimated penalty
parameters, not simultaneous confidence bands.
No composite score was calculated across tasks.

\section{Results}

\subsection{T1: SOC concentration estimation}

\begin{figure}[!ht]
\centering
\includegraphics[width=\linewidth,height=0.64\textheight,keepaspectratio]{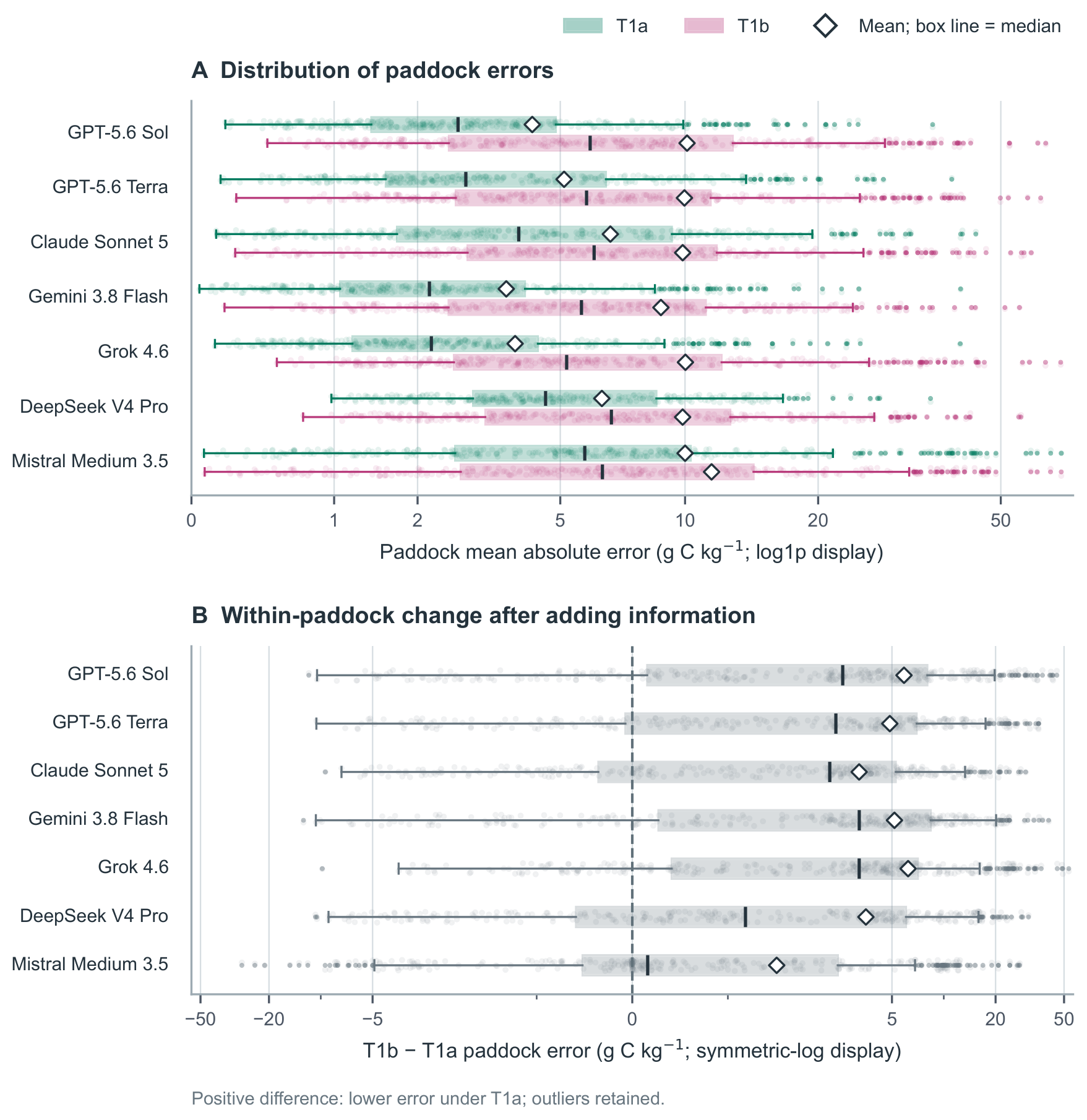}

\caption{Paddock-level SOC concentration error distributions.
(A) For each of the seven configurations, each T1a or T1b box
contains 308 paddock-level mean absolute errors, calculated by
averaging call-level absolute error over the valid finite responses
for that paddock and condition. Open diamonds show the mean of
these 308 values, which equals the reported MAE; box centre lines
show medians.
(B) Each box contains 308 within-paddock differences in mean
absolute error (T1b $-$ T1a). Positive differences indicate lower
error under T1a; open diamonds show the mean paired difference.
Boxes span Q1--Q3, whiskers reach the most extreme values within
$1.5 \times \mathrm{IQR}$, and outliers are retained.
Box statistics were computed in the original
g~C~kg$^{-1}$ units; Panel A uses a log1p display and Panel B
a symmetric-log display. No T1 confidence interval or hypothesis
test is implied. Supervised baselines are reported in
Section~3.1 and Table~S6.}
\label{fig:main3}
\end{figure}

Additional measured soil attributes and historical land use reduced MAE
for all seven configurations (Fig.~\ref{fig:main3}; Table~\ref{tab:main4}). T1b MAE was 8.78--11.51
g C kg-1 versus 3.62--10.00 g C kg-1 under T1a, with paired reductions
of 1.51--6.20 g C kg-1 (13--62\%). Gemini 3.8 Flash and Grok 4.6 had the
lowest T1a errors (3.62 and 3.82 g C kg-1), whereas Mistral Medium 3.5
had the highest (10.00 g C kg-1) and smallest reduction.

\begin{table}[!htbp]
\centering
\caption{Summary performance across the three quantitative SOC tasks}
\label{tab:main4}
\begin{tabular}{cccccc}
\toprule
\textbf{Configuration} & \textbf{T1a MAE} & \textbf{T1b MAE} & \textbf{T1b - T1a} & \textbf{T2 accuracy (95\% interval)} & \textbf{T3 accuracy (95\% interval)} \\
\midrule
GPT-5.6 Sol & 4.24 & 10.11 & +5.87 & 84.0\% (83.2--84.7\%) & 94.0\% (93.9--94.2\%) \\
GPT-5.6 Terra & 5.11 & 9.97 & +4.86 & 78.2\% (77.3--79.1\%) & 94.2\% (94.0--94.3\%) \\
Claude Sonnet 5 & 6.65 & 9.87 & +3.22 & 76.7\% (75.6--77.7\%) & 94.3\% (94.2--94.4\%) \\
Gemini 3.8 Flash & 3.62 & 8.78 & +5.17 & 89.0\% (88.4--89.7\%) & 94.6\% (94.5--94.7\%) \\
Grok 4.6 & 3.82 & 10.02 & +6.20 & 83.2\% (82.6--83.8\%) & 93.7\% (93.4--93.9\%) \\
DeepSeek V4 Pro & 6.34 & 9.87 & +3.53 & 76.1\% (74.8--77.4\%) & 93.5\% (93.2--93.8\%) \\
Mistral Medium 3.5 & 10.00 & 11.51 & +1.51 & 38.9\% (38.1--39.6\%) & 90.5\% (90.3--90.6\%) \\
\bottomrule
\end{tabular}
\par\smallskip
\begin{minipage}{\textwidth}
\footnotesize
\noindent\textit{Note.} T1 MAE and T1b - T1a are in g C kg\textsuperscript{-1}; a positive
paired difference denotes lower error under T1a. The paired differences
were calculated from unrounded paddock scores and may differ slightly
from subtraction of rounded MAEs. T2 and T3 intervals are pointwise
repeated-call 95\% intervals. 
\end{minipage}
\end{table}

Paddock-level errors remained heterogeneous under both conditions (Fig.~\ref{fig:main3}). Every configuration showed some increased errors after information
was added; Mistral Medium 3.5 had a near-zero median paired improvement.
Long upper error tails persisted despite lower average MAE.

\begin{figure}[!htbp]
\centering
\includegraphics[width=\linewidth,height=0.64\textheight,keepaspectratio]{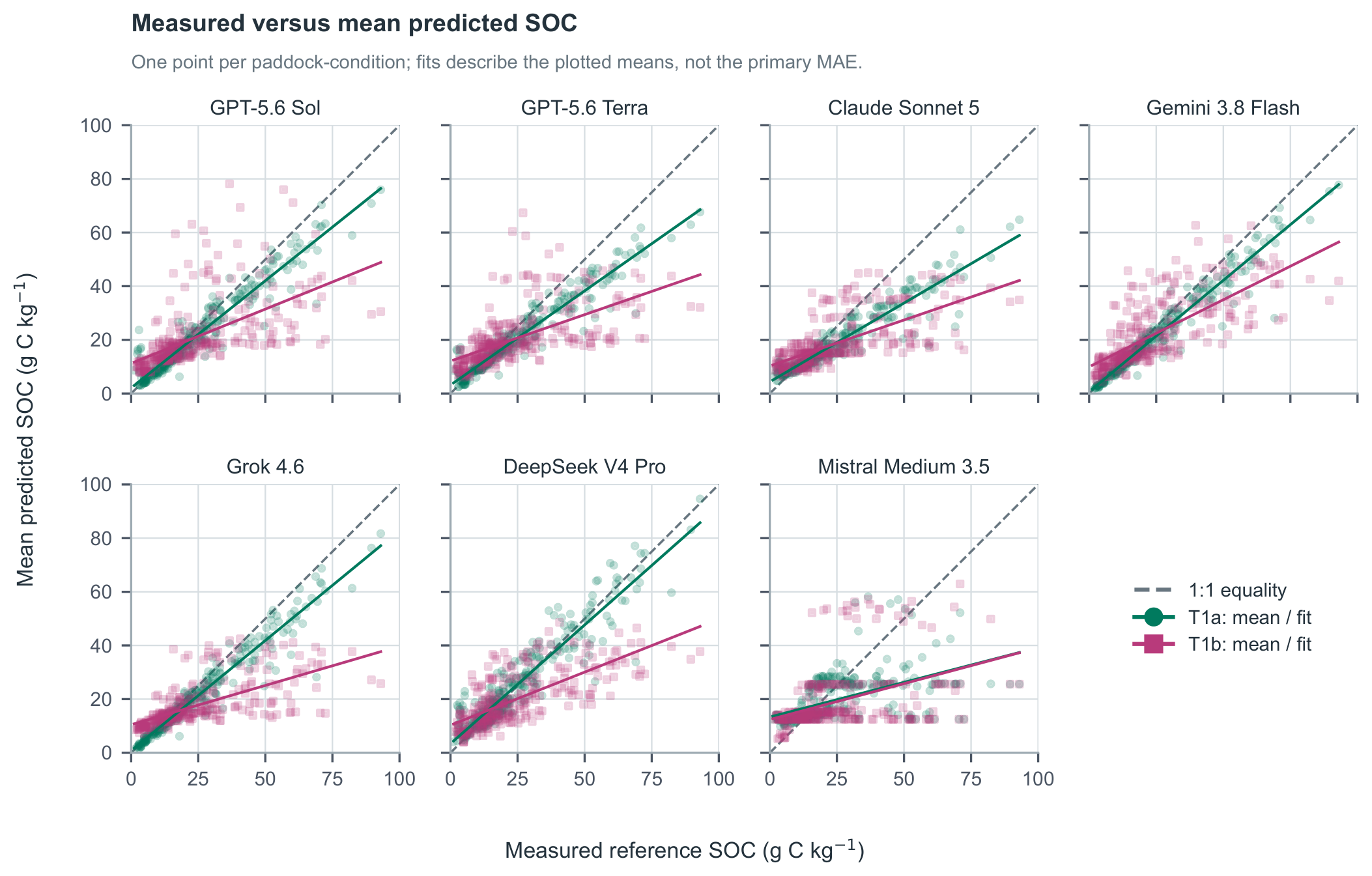}
\caption{Measured versus predicted SOC concentration. Each of
the seven facets shows the measured 0--10 cm paddock reference against
the mean valid finite prediction for that paddock under T1a (green
circles) and T1b (pink squares), with 308 paddocks per condition. The
dashed line is 1:1 equality; solid condition-specific lines are
descriptive ordinary-least-squares fits with freely estimated
intercepts, drawn only over the observed reference range. Axes are
linear, share the same scale across facets, and express SOC
concentration in g C kg\textsuperscript{-1}. A plotted point averages
valid repeated predictions; it is not an additional paddock or an MAE
observation. The fitted lines are descriptive; they do not calibrate
predictions and are not used for tests or intervals. MAE was scored from
individual responses and generally differs from the absolute error of a
mean prediction.}
\label{fig:main4}
\end{figure}

Under T1b, measured-versus-predicted fits were shallower than the 1:1
line, particularly at high reference SOC (Fig.~\ref{fig:main4}). Six configurations
showed steeper descriptive fits with T1a information, but
underprediction remained at the upper end. Mistral Medium 3.5 showed
little change. These fits were descriptive, not externally validated
calibrations.

Spatially cross-validated baselines yielded MAEs of 13.48 g C kg-1 for
B0, 18.51 for B1, 14.29 for B2$^{\prime}$ and 3.13 for B2 (Table S6).  Additional
measured predictors reduced random-forest MAE by 11.16 g C kg-1, more
than for any LLM. All T1b LLM MAEs were below B0, but these are not
like-for-like model comparisons: supervised baselines used measured
labels and spatial holdout.

\subsection{T2: SOC stock calculation}

Overall T2 accuracy ranged from 38.9\% to 89.0\% (Table~\ref{tab:main4}). Under
equation-native units without coarse fragments (U0F0), six
configurations achieved 92.8--100.0\% accuracy; Mistral Medium 3.5
achieved 58.3\% (Fig. 5A; Table S7).

\begin{figure}[!ht]
\centering
\includegraphics[width=\linewidth,height=0.64\textheight,keepaspectratio]{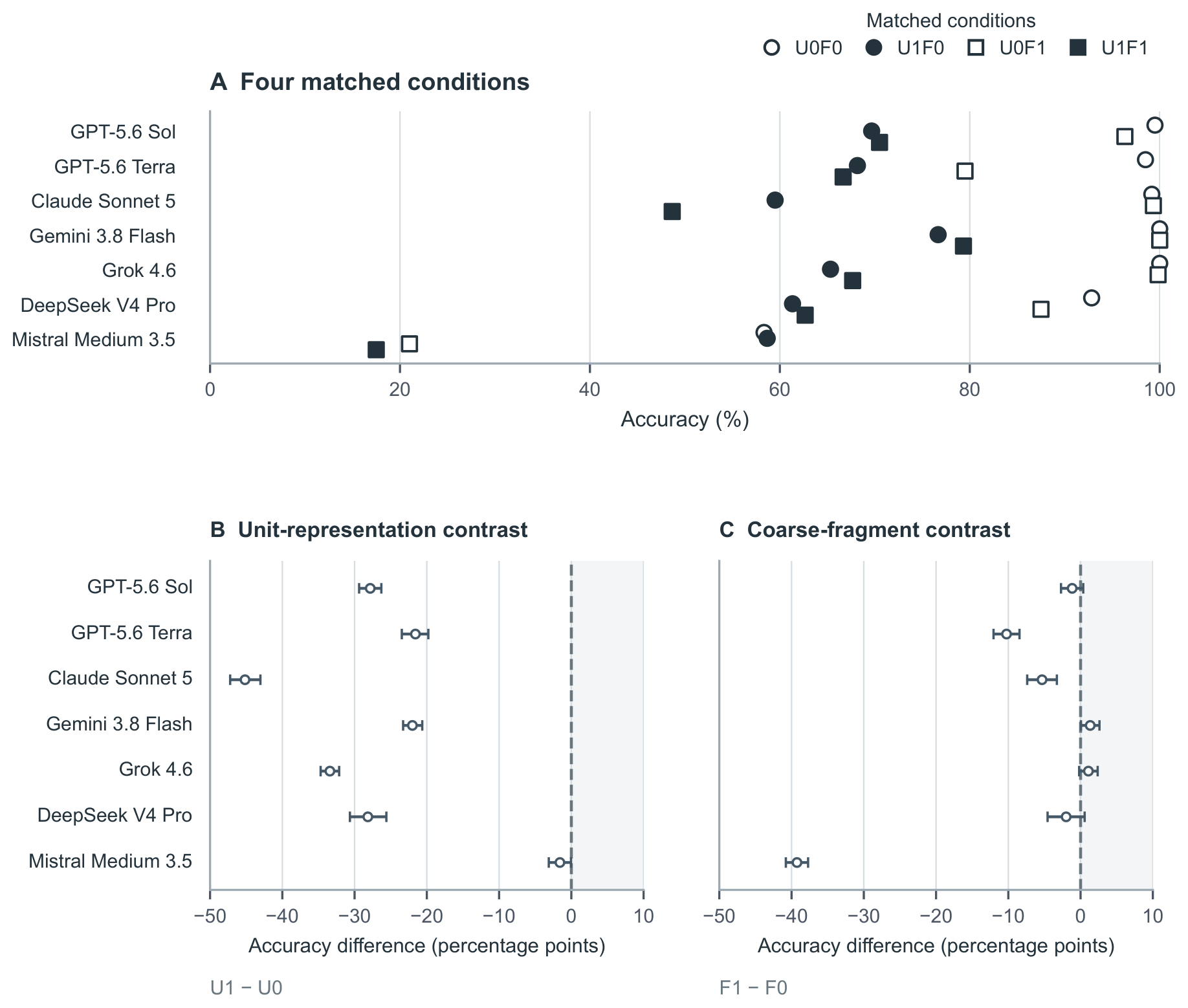}

\caption{SOC stock calculation under matched conditions.
(A) Accuracy in the four matched unit-representation (U) and
coarse-fragment (F) conditions for each configuration. Open versus
filled symbols indicate U0 versus U1; circles versus squares
indicate F0 versus F1. U1 changes one unit-representation dimension
(SOC concentration, bulk density, layer thickness, or requested
output unit) relative to U0. F1 changes the physical coarse-fragment
condition and its reference stock.
(B) Matched U1 $-$ U0 accuracy difference, averaged equally over F
and the 30 base scenarios.
(C) Matched F1 $-$ F0 accuracy difference, averaged equally over U
and base scenarios. Points are the matched contrasts and horizontal
bars their pointwise repeated-call 95\% intervals, in percentage
points; dashed lines indicate zero. Panels B and C share a
$-50$ to $+10$ percentage-point display range.
Accuracy uses the prespecified absolute-relative-error tolerance
of $\leq 2\%$, with invalid and terminally unavailable responses
scored as incorrect. U $\times$ F interaction contrasts are
reported in the Results section.}
\label{fig:main5}
\end{figure}

Unit re-expression reduced accuracy in all configurations (Fig.~\ref{fig:main5}B). The
matched U1 - U0 contrast was -45.2 to -21.6 percentage points for six
configurations and -1.6 for Mistral Medium 3.5; all pointwise 95\%
repeated-call intervals excluded zero. The F1 - F0 coarse-fragment
contrast was -5.3 to +1.3 points for five configurations, -10.3 for
GPT-5.6 Terra and -39.3 for Mistral Medium 3.5 (Fig.~\ref{fig:main5}C).

Overall accuracies of 76.1--78.2\% concealed different condition
profiles for GPT-5.6 Terra, Claude Sonnet 5 and DeepSeek V4 Pro. Claude
exceeded 99\% in both U0 conditions but scored 59.5\% and 48.7\% in U1F0
and U1F1; Terra scored 79.5\% under U0F1. Mistral ranged from 17.5\% to
58.7\% across the four conditions. The U $\times$ F interaction contrasts (Table S7)  further distinguished the condition profiles. GPT-5.6 Terra had a
positive U $\times$ F contrast of 17.5 percentage points, whereas Claude Sonnet 5
had a negative contrast of -11.0 points. The other five interactions
remained within ±7 points.

\subsection{T3: Pairwise SOC comparison}

\begin{figure}[!htbp]
\centering
\includegraphics[width=\linewidth,height=0.64\textheight,keepaspectratio]{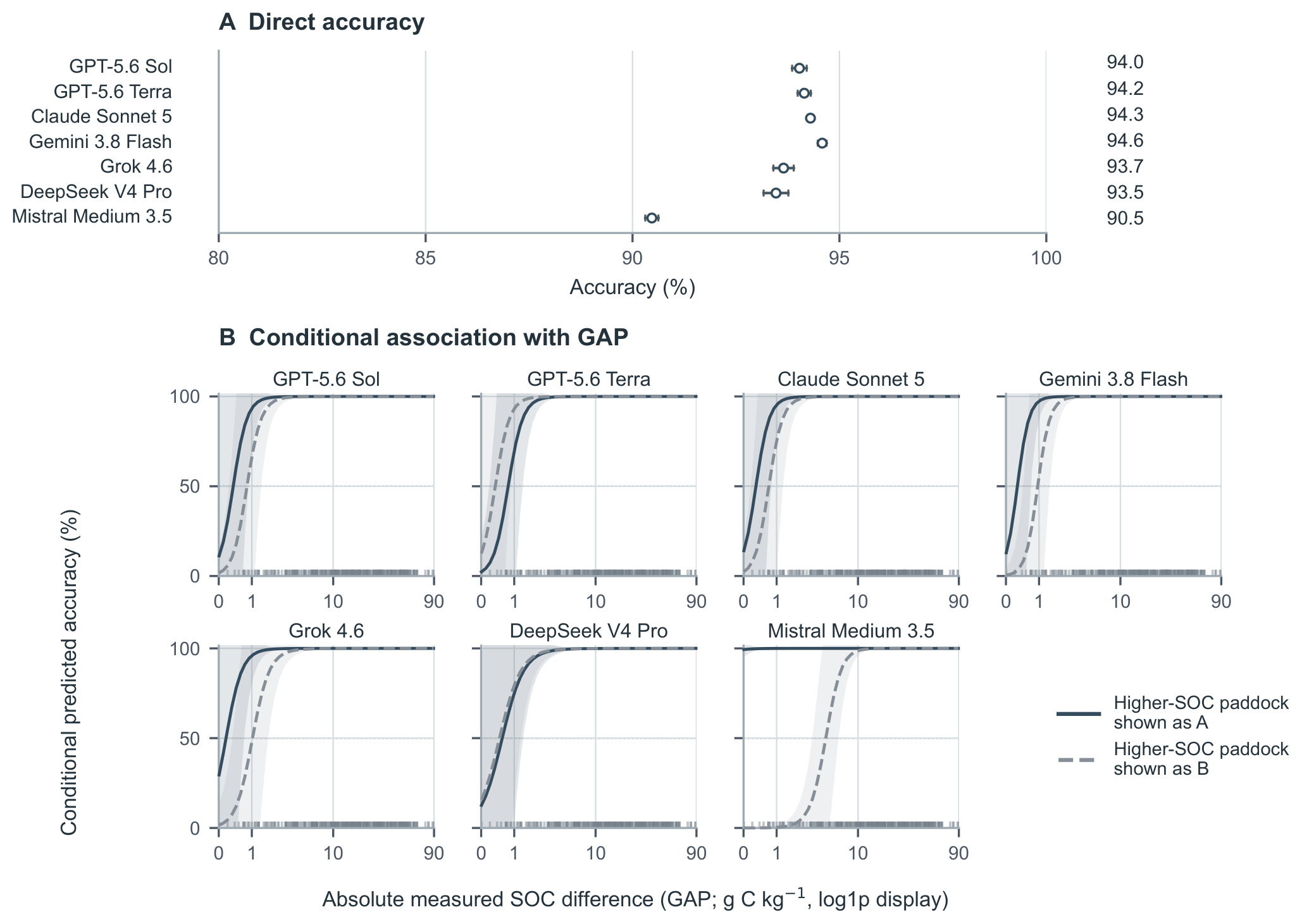}
\caption{Pairwise SOC comparison accuracy and association with
SOC separation. (A) Direct accuracy gives each of the 300 paddock pairs
equal weight and balances the two presentation orders within pair.
Points and horizontal bars are the direct accuracy estimates and
pointwise repeated-call 95\% intervals; the values at right show
one-decimal percentages. The horizontal axis in panel A is restricted to
80--100\% for display. (B) Conditional predicted accuracy from the
secondary model against the continuous absolute measured SOC difference
(GAP, g C kg\textsuperscript{-1}; log1p display), separately for the
higher-SOC paddock shown as A or B. Shading gives pointwise, not
simultaneous, 95\% Wald intervals conditional on estimated penalty
parameters. Shared-paddock, pair and configuration-by-pair random
effects are set to zero for these curves; they are not an average of the
direct scores in A. Rug marks show the 300 observed pair GAP values. The
fitted model has no GAP $\times$ position interaction on the logit scale.}
\label{fig:main6}
\end{figure}

Direct T3 accuracy ranged from 90.5\% to 94.6\% across 300 pairs (Fig.~\ref{fig:main6}A; Table~\ref{tab:main4}). Six configurations clustered at 93.5--94.6\%, compared
with 90.5\% for Mistral Medium 3.5. Eighteen of 21 prespecified
between-configuration contrasts were significant after Holm correction,
despite only a 1.1-percentage-point range among the six higher-scoring
configurations (Table S9).

Reference SOC separation strongly affected accuracy: it was 97.0--99.6\%
for 248 pairs meeting the nominal separation threshold, but 59.1--70.9\%
for the remaining 52 pairs (Table S10).  These 52 pairs represented 17\%
of the sample and accounted for most errors. The continuous-GAP model
similarly showed rising conditional accuracy with greater separation
(Fig.~\ref{fig:main6}B).

The estimated effect of presenting the higher-SOC paddock as A rather
than B was +18.1 percentage points (95\% interval 9.8--26.4) for Mistral
Medium 3.5, versus -1.6 to +3.4 for the other configurations (Table S8). 
Gemini 3.8 Flash showed a smaller positive effect (+2.7; 0.1--5.2),
whereas five intervals included zero. These were conditional model
estimates, not unadjusted differences between orders.

\subsection{Response validity, repeated-call variability and estimated cost}

Of 145,040 planned responses, 145,011 were valid (99.98\%), with at
least 99.7\% validity in each configuration--task cell (Fig. 7B; Table S5).  Fallback parsing accounted for 10.1\% of valid responses,
concentrated in Gemini 3.8 Flash (67.9\% in T1, 91.3\% in T2, 32.8\% in
T3) and Claude Sonnet 5 (86.8\% in T2). These proportions use each
cell's valid responses as the denominator. Repeated-call dispersion varied by configuration and condition (Fig. 7A; Table S11). T1 IQRs generally narrowed with T1a for several
configurations, including Gemini 3.8 Flash, but widened for DeepSeek V4
Pro. Mistral Medium 3.5 often had near-zero T1 IQRs despite large
estimation errors. T2 relative IQRs were frequently near zero but had
long upper tails, especially for Claude Sonnet 5 and Mistral Medium 3.5;
Claude had extreme item-level dispersion.

For T3, the proportion of estimable pair--order cells with nonzero
choice disagreement ranged from 0.3\% to 10.7\% across configurations
and canonical presentation orders (Supplementary Table S12).  All 300
cells in each configuration--order group were estimable. This proportion
describes cells with nonzero disagreement among valid choices, not an
error rate or the mean disagreement; canonical forward and reverse
orders do not by themselves identify the higher-SOC paddock's A/B
position. 

Estimated API cost was US\$174.64: US\$86.30 for T1, US\$29.85 for T2
and US\$58.49 for T3 (Fig. 7C; Table S13).  Configuration totals ranged
from US\$7.30 (DeepSeek V4 Pro) to US\$80.92 (Grok 4.6), an
approximately elevenfold difference. A separate US\$1.55 reservation for
uninterpretable token usage was excluded. These were standardised
provider-rate estimates, not invoiced costs.

\begin{figure}[!ht]
\centering
\includegraphics[width=\linewidth,height=0.64\textheight,keepaspectratio]{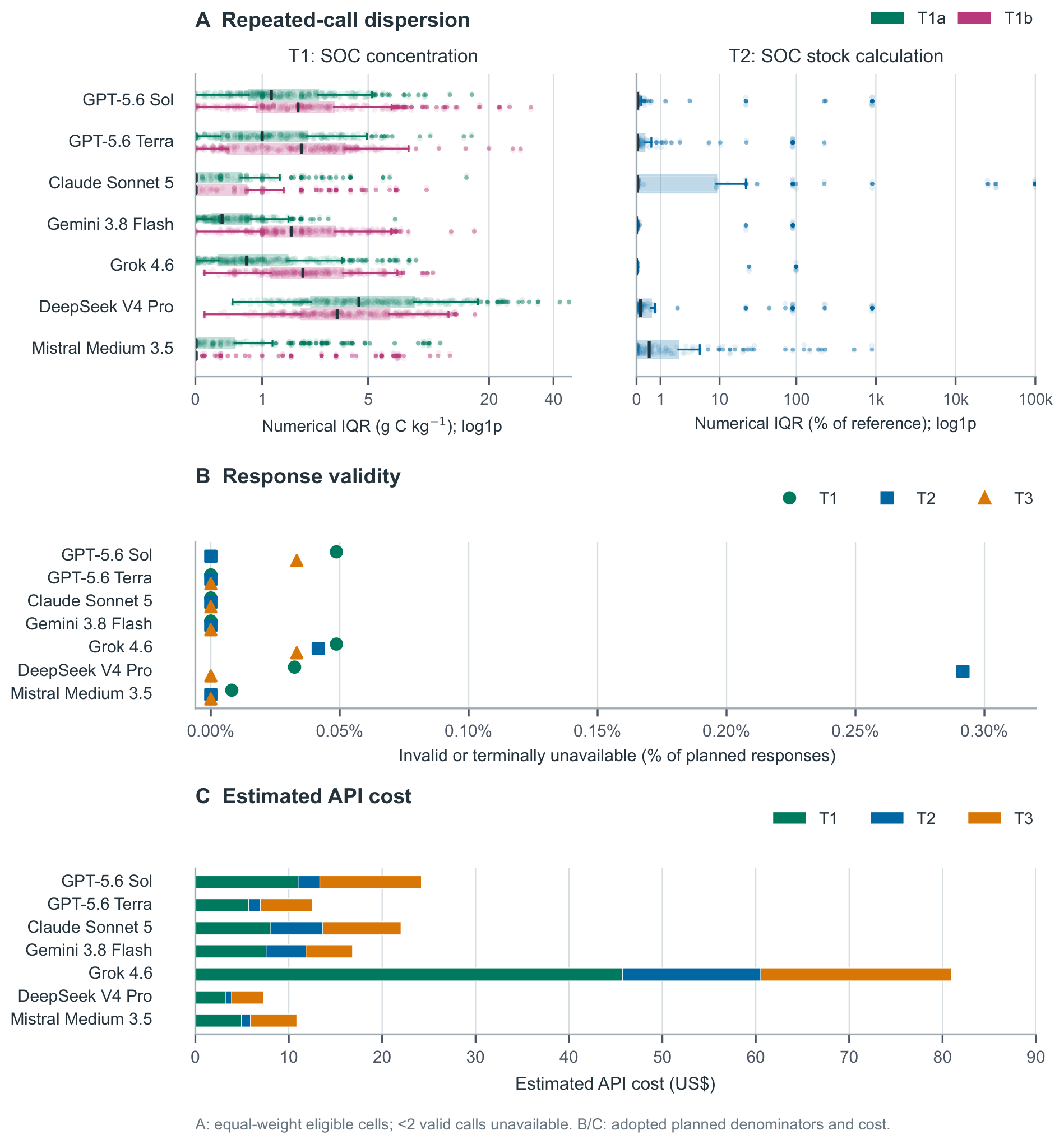}
\caption{Repeated-call dispersion, response validity and
estimated API cost. (A) Post-collection descriptive summary of
within-cell numerical-answer dispersion for T1 and T2. T1 shows the
interquartile range (IQR) of valid finite SOC predictions within each
configuration--paddock--condition cell (g C kg\textsuperscript{-1}),
separately for T1a and T1b. T2 shows 100  $\times$  IQR of valid finite stock
predictions divided by the absolute reference stock in the same
requested output unit, within each configuration--item cell. Each
estimable cell contributes once; at least two valid responses are
required. Zeros and the complete observed upper tail are retained, and
log1p affects only display, not box statistics. (B) Invalid returned
outputs plus responses that remained unavailable after operational
failures, as percentages of planned responses; both are counted
separately in Table S5.  This is response validity, not accuracy. (C)
Estimated API cost (US\$) by task and configuration, based on
provider-reported usage rather than invoiced charges. The separate
US\$1.55 reservation for attempts without interpretable usage is
excluded. T3 within-order choice disagreement is reported in Supplementary Table S12.}
\label{fig:main7}
\end{figure}

\section{Discussion}

\subsection{Measured inputs improve SOC indicator agreement but do not eliminate
bias}

The improvement with T1a information shows that LLM estimates respond to
soil observations, not that they independently measure SOC. The added
predictor block included bulk density, total nitrogen, pH, electrical
conductivity, coarse fragments and historical land use. Supplying
comparable predictors reduced the spatially cross-validated
random-forest MAE by 11.16 g C kg-1, indicating substantial information
content. Carbon--nitrogen coupling makes nitrogen a plausible
contributor \cite{main_ref_027}, but individual effects cannot be
separated in this blockwise design. For monitoring, model-derived SOC
must therefore be distinguished from laboratory observation and
interpreted with the provenance and limitations of its inputs.

Local errors and high-SOC underprediction matter beyond average MAE.
Where measured SOC supports locally defined status classes, compressed
high-end predictions could mischaracterise carbon-rich paddocks or
reduce apparent contrasts. We did not calculate class changes or
management outcomes; the concern is an interpretive risk, not an
observed decision error. Soil- and climate-dependent SOC reference
systems further limit universal thresholds \cite{main_ref_003,main_ref_015}. Concentration-dependent bias should therefore be
checked before applying estimates to ecological assessment, rather than
translating these results into a general soil-health classification. Sparse-input performance also has limits. The same-state median and
sparse-input random forest did not outperform the global median under
spatial holdout, which is not interchangeable with map-accuracy
evaluation \cite{main_ref_051}. LLMs may encode geographic information
from pretraining \cite{main_ref_033,main_ref_049}, and exposure
to publicly available SOC-M or related soil-carbon data cannot be
excluded \cite{main_ref_002,main_ref_024,main_ref_012}. These results do not establish generalisation to unseen regions
or the mechanism underlying location-based estimates.

SOC is only one dimension of soil condition: physical, chemical and
biological indicators are needed to assess wider soil functions \cite{main_ref_010,main_ref_022,main_ref_056}. Unverified SOC
numbers cannot provide a secure basis for site-specific ecological or
management claims.

\subsection{Equivalent units expose a failure of stock-indicator consistency}

Six configurations performed near ceiling on stock calculations in
native units but lost approximately 22--45 percentage points of accuracy
under equivalent unit re-expression. Unlike concentration, areal SOC
stock depends on depth, bulk density and mass-basis conventions. These
physical inputs can alter true calculated stock; unit notation alone
cannot. The matched U contrast therefore tests whether the same carbon
quantity survives conversion, and shows that supplying an equation did
not ensure consistency. Similar sensitivity to numerical substitutions
has been reported in other LLM tests \cite{main_ref_037}, although
T2 specifically involved physical units.

Adding coarse fragments, unlike changing units, alters the reference
stock. Correct use of the supplied dry-mass fraction remained necessary,
given the sensitivity of stocks to bulk-density and fragment treatment
\cite{main_ref_018,main_ref_041}. The smaller fragment
penalty for most configurations does not establish that fragment
handling is intrinsically easier; the two contrasts test different
operations.

For stock reporting, the input mass basis, unit conversions, fragment
correction and arithmetic should be independently auditable. A
deterministic implementation of the stated equation provides the
appropriate calculation reference; tool-enabled comparisons and
individual conversion types remain useful targets for future validation.

This distinction matters for monitoring, reporting and verification
(MRV): an unchecked calculation error could enter a reported stock or
apparent change even when measured soil properties were unchanged.
Transparent sampling and stock-accounting procedures remain essential
\cite{main_ref_034,main_ref_045}. Because T2 was synthetic,
it did not measure farm-level changes, audit an MRV programme or
establish the effectiveness of proposed safeguards.

\subsection{SOC discrimination requires separation- and order-aware evaluation}

High mean T3 accuracy concealed weak discrimination among closely
matched paddocks. Exploratory ranking can be informative only when sites
share comparable sampling depth, soil context and reference basis.
Otherwise an incorrect ordering could mislead interpretation or
prioritisation, although no such decisions were measured. The nominal
GAP split identifies difficult comparisons within this sample, not an
ecological or administrative action threshold.

Small GAP also does not imply that measured reference orderings were
wrong. Reference SOC estimates have sampling and laboratory uncertainty,
and field heterogeneity may exceed the changes sought in monitoring
\cite{main_ref_047}. The 10\% separation convention did not quantify
paddock-specific uncertainty. T3 compared different paddocks, not
changes over time; monitoring interpretation requires an appropriate
reference population and assessment objective \cite{main_ref_003,main_ref_058}. Reversing A/B order preserves the reference ordering yet affected some
LLM responses, most notably Mistral Medium 3.5. Such position effects
have been observed in other language-model evaluations \cite{main_ref_046,main_ref_057}. Balancing order reveals the bias but does not
remove it; no GAP-by-position interaction was fitted. Applications
requiring comparisons should test both order and performance near
relevant SOC separations.

Relative SOC status is also distinct from attainable carbon gain. Lower
current SOC does not necessarily imply greater management potential:
mineralogy, climate, historical disturbance and feasible practices
constrain stocks. No such intervention potential was evaluated in T3, so
pairwise rankings should not be read as management prescriptions.

\subsection{Output validity and response dispersion are distinct from indicator
quality}

Repeated-call consistency is not evidence of correctness. Mistral Medium
frequently produced near-zero T1 IQR despite high error, while T2
contained extreme item-level answer ranges (Fig.~\ref{fig:main7}A). IQR measures
dispersion among outputs, not reference uncertainty; identical-looking
repeated results still require checking against measurement or a
deterministic calculation \cite{main_ref_001}. Likewise, 99.98\% scoreability is a response-format result, not
indicator accuracy. Fallback parsing was common in selected model--task
cells, underscoring the need to distinguish strict compliance from
answers accepted by permissive rules.

The approximately elevenfold cost difference reflects recorded provider
prices and token accounting, not intrinsic computational efficiency or
environmental impact. Validity, error, repeat dispersion and estimated
cost should remain separate reporting dimensions.

\subsection{Implications for ecological-indicator monitoring and management}

The task-specific findings support a proposed quality-control sequence
(Table~\ref{tab:main5}): define whether the target is concentration, areal stock or
relative ordering; record sampling depth, measurement provenance and
mass basis; then check the relevant operation against independent
measurements (T1), deterministic unit-consistent calculations (T2), or
reference separation and order reversal (T3). These checks follow
observed failure modes but were not evaluated as a deployed management
system. The benchmark statistics describe model reliability, not
additional ecological state indicators \cite{main_ref_011,main_ref_019}.

\begin{table}[!htbp]
\centering
\caption{Practical interpretation of the benchmark for soil organic carbon indicator workflows.}
\label{tab:main5}
\begingroup
\setlength{\tabcolsep}{3pt}
\begin{tabular}{@{}
  >{\raggedright\arraybackslash}p{0.227\linewidth}
  >{\raggedright\arraybackslash}p{0.228\linewidth}
  >{\raggedright\arraybackslash}p{0.227\linewidth}
  >{\raggedright\arraybackslash}p{0.228\linewidth}@{}}
\toprule
\textbf{Indicator operation} & \textbf{Benchmark evidence} & \textbf{Interpretive limitation} & \textbf{Minimum check before applied use} \\
\midrule
SOC concentration (T1) & Added measurements improved average MAE;
high-SOC underprediction and local errors remained. & Estimated SOC may
mischaracterise carbon-rich sites; no soil-health classes or management
effects were tested. & Verify against independent 0--10 cm measurements,
suitable local references and concentration-dependent bias. \\
SOC stock (T2) & Equivalent-unit re-expression reduced correctness under
a supplied stock equation. & A unit-induced stock difference is an
arithmetic artefact, not a change in soil carbon storage. & Document
concentration mass basis, depth, bulk density and fragment convention;
recalculate deterministically. \\
Paddock comparison (T3) & Accuracy fell for close SOC pairs; some A/B
position effects were observed. & Relative SOC ranking does not
establish the potential for carbon gain or a causal management response.
& Compare like soil contexts; quantify measured uncertainty; verify both
presentation orders. \\
Operational quality & Almost all responses were parseable, although
repeated-call dispersion and estimated costs varied. & Formatting,
reproducibility and price are not substitutes for ecological or
numerical validity. & Audit strict validity, input provenance,
correctness, repeat variation and costs separately. \\
\bottomrule
\end{tabular}
\endgroup
\end{table}

Appropriate safeguards depend on use. For education or early-stage
extension, an explicitly provisional estimate may prompt discussion
about required measurements. For monitoring and land-management
prioritisation, biased concentration estimates or unstable ordering
warrant local references and independent review. Contractual carbon
reporting additionally requires traceable stock calculations and
established verification procedures \cite{main_ref_034,main_ref_045}. These are proposed decision pathways, not management effects
observed in this study.

Outcome-focused ecological monitoring depends on reliable links from
measured soil properties to interpretation and management \cite{main_ref_053}. Our study tests the numerical link for SOC, not ecosystem
services or ecological responses; broader soil-health inference requires
complementary indicators and appropriate reference systems \cite{main_ref_022,main_ref_058}.

\subsection{Scope, limitations and future validation}

The 308 empirical references cover sampled 0--10 cm areas, not whole
paddocks or deeper profiles, and include sampling and analytical
uncertainty. T2 comprised 30 synthetic scenarios; T3 reused the same
paddock pool. Repeated-call intervals do not represent performance
uncertainty for new farms or scenarios, and collection used one prompt
and reasoning setting per configuration within a short period.
Provider-side model versions were not independently verified;
descriptive fits and dispersion summaries were prepared after
collection. The study did not evaluate ecological thresholds,
intervention outcomes, decision losses or tool-augmented deployment.
Future validation should use independent sites, quantify reference
uncertainty and prespecify tolerances linked to particular monitoring or
management purposes.

\section{Conclusions}

Across three SOC tasks, LLM reliability depended on both the
environmental information supplied and the representation of the
problem. Added soil attributes reduced concentration error, but high-SOC
underprediction and local failures remained. Stock calculations were
vulnerable to physically equivalent changes in units, while comparisons
were less accurate for similar paddocks and sometimes sensitive to
presentation order. Near-universal parseability and low repeated-call
dispersion did not guarantee correctness. The ecological-indicator
contribution is to distinguish errors in estimated soil condition,
derived stock and relative site ordering. These failure modes matter for
interpretation, but their effects on management decisions were not
tested. Before applied use, SOC estimates require measured validation,
stock figures require auditable unit handling, and rankings require
separation- and uncertainty-aware review. The benchmark provides a
direct-response reference for future field-oriented, tool-augmented
evaluation---not evidence of soil-health improvement or carbon
sequestration.

\section*{Data availability and reproducibility}

The source Soil Organic Carbon Monitoring (SOC-M) dataset (version 1) is
publicly available through the CSIRO Data Access Portal
(https://doi.org/10.25919/z378-4286). Benchmark item definitions, prompt
templates, scoring conventions and collection settings are described in
the Methods section of this manuscript.

\section*{Funding}

This research was funded by the National Foundation for Australia-China
Relations {[}NFACR240313{]} and the Australian Research Council
{[}DP230101671{]}.

\section*{Declaration of competing interests}

No.

\section*{CRediT author contributions}

Yingxuan Cheng: Writing -- review \& editing, Writing -- original draft,
Visualization, Software, Methodology, Formal analysis,
Conceptualization. Tong Li and Jinran Wu: Supervision; Writing -- review
\& editing, Visualization, Software, Methodology, Formal analysis,
Conceptualization. Yayong Li, Yuchuan Fan, You-Gan Wang, and Geoffrey J. McLachlan:
Writing -- review \& editing.

\section*{Declaration of generative AI and AI-assisted technologies}

A large language model (LLM) was used to assist with the language
editing of this manuscript. The authors take full responsibility for the
content and interpretation presented in this work.

\section*{Acknowledgements}

NA.

\printbibliography[title={References}]



\section*{Supplementary material}

# Supplementary Information

Reliability of large language models for soil organic carbon indicators: An Australian benchmark of estimation, stock calculation and pairwise comparison

## Supplementary Methods

### M1 Data provenance and reference construction

The empirical dataset was version 1 of the CSIRO Soil Organic Carbon Monitoring (SOC-M) release (csiro:69217v1; DOI: 10.25919/z378-4286). Of 3,644 records in SOCM_AllAnalytical.csv, 1,213 had numerical upper and lower depths of 0 and 10 cm, respectively, representing 308 distinct SCaRP_ID values. Each paddock's reference SOC concentration was the unweighted mean of its retained OC_ODSoil values. This oven-dry-corrected variable is reported in mg C $g^{-1}$ soil, numerically equivalent to g C $kg^{-1}$. Records within a paddock were not treated as independent paddocks. Table S1 gives record counts and source-site classifications (Karunaratne et al., 2026; Pasut et al., 2026).

No analytical-method exclusions or inter-method conversions were applied. Although the reference measurements were classified within a dry-combustion method family, the available source dictionary and methodology report do not establish the exact method or fine-earth versus whole-soil mass denominator for each OC_ODSoil record (Farrell et al., 2025). ManageClass_2012 describes historical land use, not necessarily management at the later sampling date.

State, coordinates, GRDC region and historical land use were linked by SCaRP_ID from SOCM_FarmDataManagemement.csv; depths and soil properties came from the retained analytical records. Each paddock contributed one reference value and one input-field block. Figure 1 in the main text uses source-reported locations and ABS ASGS Edition 3 (2021) boundaries, projected to Australian Albers. WGS84 was assumed for the source coordinates but was not verified for every record. Table S2 lists the input fields and their missingness. Missing prompt fields were omitted, whereas random-forest predictor imputation was fitted within training folds.

### M2 Model requests and collection protocol

The seven API configurations, provider-native request fields and output limits are listed in Table S3. Each planned response was an independent, stateless, closed-book call without conversation history, retrieval, executable tools or reference answers. Temperature, top-p and seed were unspecified. Reasoning settings were none, disabled or low, as applicable; these controls are not equivalent across providers. Each request specified an output limit of 1,024 under its provider's token-accounting convention, which did not necessarily correspond to 1,024 visible tokens.

Calls followed a serial, interleaved schedule: 616 T1 paddock–condition items, 120 T2 items and 300 T3 pairs, with 20 planned responses per configuration–item or pair. Across seven configurations, the planned totals were 86,240 (T1), 16,800 (T2) and 42,000 (T3), or 145,040 overall. Up to four physical attempts per planned response were allowed for operational failures; retries preserved the same logical response position. Incorrect or unparseable answers were not retried. The collection audit recorded 145,117 attempt intents, 145,016 returned outputs, five invalid outputs and 24 terminally unavailable responses (Table S5).

### M3 T1 prompts and supervised baselines

T1b supplied state, source-precision coordinates, depth (0–10 cm) and GRDC region when available. T1a added historical 2012 land use, whole-soil bulk density, coarse-fragment dry-mass fraction, total nitrogen, electrical conductivity and $CaCl_2$ pH. The sole missing GRDC value was omitted in both conditions. The common T1 system prompt was:

> You are given structured information about a soil observation.
> Estimate the soil organic carbon (SOC) concentration.
> Report the value in g/kg.
> Respond with a single JSON object and nothing else.

User prompts used the following field order, omitting unavailable fields; the six added property fields appeared only in T1a. Braced expressions below indicate substitution slots, not literal model inputs. Numerical precision matched the archived rendered items.

> State: {state}
> Latitude: {latitude} decimal degrees
> Longitude: {longitude} decimal degrees
> Upper soil depth: {upper depth} cm

Lower soil depth: {lower depth} cm
GRDC region code (categorical): {GRDC region}
Historical/baseline 2012 land-use class: {land-use class}
Whole-soil bulk density: {bulk density} g/cm3
Coarse-fragment dry-mass fraction (>=2 mm; 0-1): {fraction} g/g
Total nitrogen (oven-dry soil): {total nitrogen} mg N/g soil (= g N/kg)
Electrical conductivity (1:5 soil:water): {conductivity} µS/cm
pH (0.01 M CaCl2): {pH}

Return: {"prediction": <number in g/kg>}

Four label-informed baselines used the same 308 paddocks and five spatial folds (79, 79, 33, 21 and 96 paddocks; seed 2026092902). Approximate planar-kilometre coordinates were grouped by five-cluster k-means without SOC labels. B0 predicted the training-fold SOC median; B1 used the same-state training median, with B0 as fallback. B2 and B2′ were random forests with T1a and T1b inputs, respectively (Breiman, 2001). Both forests used 500 trees, bootstrapping, squared-error splits, min_samples_leaf=2, max_features=1.0 and seed 2026092904, without response-guided tuning. Numerical medians, categorical modes and category levels were fitted within each training fold, with unseen holdout categories assigned a dedicated level. Table S6 reports MAEs. These label-trained baselines are contextual rather than directly comparable closed-book models.

**M4 T2 scenarios, unit variants and scoring**

Thirty synthetic base scenarios crossed unit representation (U0/U1) with coarse fragments (F0/F1), yielding 120 matched items (seed 2026092903). Generator ranges were SOC 1–95 g C $\text{kg}^{-1}$, whole-soil bulk density 0.6–2.1 g $\text{cm}^{-3}$, layer thickness 5, 10, 15, 20 or 30 cm, and F1 coarse-fragment dry-mass fraction 0.08–0.40. SOC and bulk density were displayed to four significant figures, thickness in its specified unit, and fragments to one decimal percent. F0 specified 0.0% fragments; F1 specified a nonzero fraction and therefore changed reference stock. U1 changed one unit-representation dimension per base scenario, consistently across F0 and F1 (Table S4). The 120 displayed scenarios and reference values are provided in the separate machine-readable T2 CSV.

The supplied stock convention was $S = C \times BD \times d \times (1 - CF) \times 0.1$, with C in g C $\text{kg}^{-1}$, whole-soil BD in g $\text{cm}^{-3}$, d in cm, CF as a dry-mass fraction, and S in t C $\text{ha}^{-1}$. Displayed fragment percentages were divided by 100. The convention uses whole-soil bulk density and a fragment dry-mass correction; it does not independently establish the SOC mass basis required for a physical fine-earth-stock interpretation. References were recalculated from displayed values after rounding and in the requested output unit. The T2 system prompt was:

You are given the quantities required to calculate a soil organic carbon stock.
Bulk density is whole-soil bulk density.
Coarse-fragment content is the dry-mass percentage of material >= 2 mm.
Use the following stock-calculation convention: SOC × whole-soil bulk density × layer thickness × (1 - coarse-fragment dry-mass fraction) × 0.1 gives t C/ha.
Calculate the SOC stock for the described soil layer.
Report the value in {requested output unit}.
Respond with a single JSON object and nothing else.
Soil organic carbon concentration: {displayed SOC} {displayed SOC unit}
Whole-soil bulk density: {displayed bulk density} {displayed bulk-density unit}
Layer thickness: {displayed thickness} {displayed thickness unit}
Coarse-fragment content (>= 2 mm, dry mass): {displayed percentage} %

Return: {"prediction": <number in {requested output unit}>}

Finite responses within 2% absolute relative error of the deterministic displayed-value reference were correct; invalid and terminally unavailable responses were incorrect. Overall accuracy equally weighted the 20 planned binary responses per item and the 120 items. Within each base, U1 − U0 was averaged across fragment conditions, F1 − F0 across unit conditions, and U×F was $(p_{11} - p_{01}) - (p_{10} - p_{00})$; contrasts were then averaged across 30 bases. Table S7 presents accuracies, contrasts and pointwise 95% intervals; Figure S1 shows the interaction.

**M5 T3 pairs and secondary model**

From 47,278 eligible unordered paddock pairs, 300 were sampled without replacement using Python 3.14.3 random.Random (seed 2026092803), without stratification by SOC separation (GAP). Each paddock was described by its available T1a fields. For each configuration and pair, ten presentations placed the higher-reference-SOC paddock at A and ten at B. 'Forward' and 'reverse' denote canonical order, not the correct answer's position. The common T3 system prompt was:

```
You are given two soil observations, A and B.
Each observation is described using the available fields; unavailable fields are omitted.
Determine which observation has the higher soil organic carbon concentration.
Respond with a single JSON object and nothing else.
Observation A:
{available T1a-style field block for paddock A}

Observation B:
{available T1a-style field block for paddock B}

Return: {"prediction": "A"} or {"prediction": "B"}
```

The secondary analysis fitted 4,200 grouped-binomial cells (ten responses per configuration × pair × correct-answer position) using a logit link and three natural-cubic-spline terms (ns1–ns3) for log(1 + GAP), where GAP is the absolute reference-SOC difference (g C $kg^{-1}$). Boundary knots on the transformed scale were 0.006843268335663492 and 4.500598051049834; internal knots were 2.3535087659030087 and 3.2549810677995037. The fixed-effect formula was model_code × (ns1 + ns2 + ns3 + correct_position), comprising 35 columns with no GAP × position interaction. Penalised random-effect terms represented shared paddocks (Zsite, 265 columns), pairs (Zpair, 300) and configuration-by-pair combinations (Zmp, 2,100); each pair loaded its two paddocks at weight $1/\sqrt{2}$. The full 2,700-column penalised system was fitted by Schur elimination under Laplace-REML. Final penalties were 0.020725558758481682 (Zsite), 0.02758502575444963 (Zpair) and 0.06503432748159946 (Zmp). Fit diagnostics and a code hash were archived; original NumPy and SciPy package versions were not recorded.

Conditional GAP curves set random effects to zero and used the fitted fixed coefficients. Pointwise 95% Wald intervals incorporate fixed-effect conditional covariance but condition on the estimated penalties; they are neither simultaneous bands nor averages of direct accuracy. The conditional A − B position contrast averages predicted-probability differences over the 300 observed GAP values (Table S8). Table S9 reports 21 pairwise configuration differences with Holm-adjusted P values. The prespecified nominal separation rule, $|C1 - C2| \geq 2\sqrt{[(0.10C1)^2 + (0.10C2)^2]}$, was met by 248 pairs; 52 did not meet it (Table S10). The 10% assumption is a nominal comparison convention, not a measured reference-uncertainty estimate, and did not affect pair selection or the continuous-GAP model.

**M6 Parsing, aggregation and repeated-call uncertainty**

The strict parser required one task-specific JSON object containing a numerical estimate (T1/T2) or A/B choice (T3). Predefined fallbacks allowed embedded JSON, a JSON-string number, a lone number or a lone A/B label, as appropriate; references were never used to repair responses. Invalid returned output and terminal operational failure were recorded separately (Table S5). T1 absolute errors were calculated on valid finite calls, then averaged within paddock and equally across paddocks; T1b − T1a was paired within paddock when both conditions were estimable. Finite negative or extreme predictions remained in the analysis. For T2/T3, invalid and terminally unavailable positions counted as incorrect among planned responses.

For T2, 10,000 bootstrap replicates (seed 2026100201) resampled 20 binary scores within each configuration × base × U × F cell. For T3, 10,000 replicates (seed 2026092943) resampled ten binary scores within each configuration × pair × correct-position cell. All benchmark items and equal weights were retained; repeat indices were not matched across cells. Deviations from each observed estimate were rescaled by $\sqrt{[n/(n-1)]}$ ($n = 20$ for T2; $n = 10$ for T3), and pointwise 95% intervals used the 2.5th and 97.5th percentiles (Efron and Tibshirani, 1993; Hesterberg, 2015). These intervals describe repeated-call variation on fixed items, not uncertainty for new sites, scenarios or pairs. No T1 confidence intervals or tests were used. For T3 configuration comparisons, the fixed-bank variance estimate was $V_m = \Sigma_c\, p_{mc}(1 - p_{mc})/9 \div 600^2$, where $p_{mc}$ is the observed success proportion in pair–position cell c ($n = 10$). Contrast standard errors were $\sqrt{(V_a + V_b)}$, without cross-configuration covariance; two-sided normal P values were $2\Phi(-|\Delta|/SE)$. Holm adjustment covered 21 comparisons; unavailable tests were assigned P = 1 for adjustment but

reported as unavailable (Holm, 1979). Zero-variance cells retained point estimates but had unavailable repeated-call intervals; pointwise endpoints were not Holm-adjusted.

Theoretical U×F interaction bounds are −2 to 2 on the proportion scale. The archived interval routine clipped contrast endpoints to −1 to 1; saved unrounded endpoints show that this did not affect any of the seven interaction intervals. No estimates or intervals were recalculated for this supplement.

**M7 Post-collection descriptive displays**

Main Figure 3 displays paddock-level mean absolute errors and paired T1b − T1a differences; diamonds mark means and box statistics use the original g C $kg^{-1}$ scale. The log1p and symmetric-log axes affect presentation only. Main Figure 4 plots each paddock's mean valid prediction against its measured reference, with descriptive ordinary-least-squares fits; these means were not used to calculate call-level MAE.

Main Figure 7A summarises within-cell dispersion using NumPy linear-interpolation quartiles. T1 uses the IQR of valid finite SOC estimates (g C $kg^{-1}$) by configuration × paddock × condition; T2 uses 100 × IQR/|reference stock| (%) by configuration × item, with matching output units. Valid but incorrect predictions remain included. Cells with fewer than two valid answers are unavailable, not zero. Each estimable cell has equal weight; zero IQRs and extreme values are retained (Table S11). Log1p transforms apply only to the display.

T3 choice disagreement was calculated within each configuration × pair × canonical-order cell as 100 × [1 − max(nA, nB)/nvalid] among valid choices. Cells with fewer than two valid responses were unavailable; zero disagreement indicates identical choices, not correctness. Table S12 reports zero and nonzero disagreement. These post-collection analyses are descriptive and were not combined into a stability score; archived sample-standard-deviation summaries were not substituted for IQR results.

**M8 Token accounting and estimated cost**

Provider usage was normalised to avoid double counting: Gemini candidate and thought tokens were combined once, while nested reasoning tokens already included in output totals were not added again. The recorded 29 September 2026 input/output rates were applied without separate cached-input discounts. All attempts with interpretable usage, including retries, contributed to estimated cost; unknown-usage attempts retained a separate reservation rather than zero cost. Tables S13 and S14 report usage, task-level costs, rates and price sources. These are standardised estimates, not invoices or comparable measures of internal computation.

**Supplementary Tables**

**Table S1. Retained 0–10 cm analytical records per paddock**

| Site class | Retained rows per paddock | Paddocks | Analytical rows |
|---|---|---|---|
| Standard | 1 | 205 | 205 |
| Standard | 2 | 2 | 4 |
| Detailed | 9 | 6 | 54 |
| Detailed | 10 | 95 | 950 |
| Total | — | 308 | 1,213 |

**Note.** Records are analytical observations, not independent paddocks. Standard/Detailed denotes the source Site classification.

**Table S2. T1 model-visible fields and paddock-level missingness**

| Condition | Prompt field | Source column | Source file | Missing paddocks |
|---|---|---|---|---|
| T1a/T1b | State | State | FarmData | 0 |
| T1a/T1b | Latitude | Latitude | FarmData | 0 |
| T1a/T1b | Longitude | Longitude | FarmData | 0 |
| T1a/T1b | Upper soil depth | u_depth | Analytical | 0 |
| T1a/T1b | Lower soil depth | l_depth | Analytical | 0 |
| T1a/T1b | GRDC region code (categorical) | GRDC_Regions | FarmData | 1 |
| T1a only | Historical/baseline 2012 land-use class | ManageClass_2012 | FarmData | 0 |
| T1a only | Whole-soil bulk density | BD | Analytical | 0 |
| T1a only | Coarse-fragment dry-mass fraction (>=2 mm; 0-1) | Pgrav | Analytical | 0 |
| T1a only | Total nitrogen (oven-dry soil) | TN_ODSoil | Analytical | 0 |
| T1a only | Electrical conductivity (1:5 soil:water) | EC | Analytical | 0 |
| T1a only | pH (0.01 M CaCl2) | pH(CaCl2) | Analytical | 0 |

**Note.** FarmData = SOCM_FarmDataManagemement.csv (source spelling retained); Analytical = SOCM_AllAnalytical.csv. Missing prompt fields were omitted; T3 used available T1a-style fields.

**Table S3. Provider-specific request fields for seven configurations**

**Panel A. Fields sent**

| Configuration | Provider | Interface | Requested ID | Reasoning field / value | Output-limit field / value |
|---|---|---|---|---|---|
| GPT-5.6 Sol | OpenAI | Chat Completions | gpt-5.6-sol | reasoning_effort: none | max_completion_tokens: 1,024 |
| GPT-5.6 Terra | OpenAI | Chat Completions | gpt-5.6-terra | reasoning_effort: none | max_completion_tokens: 1,024 |
| Claude Sonnet 5 | Anthropic | Messages | claude-sonnet-5 | thinking.type: disabled | max_tokens: 1,024 |
| Gemini 3.8 Flash | Google | Generate Content | gemini-3.8-flash | thinkingConfig.thinkingLevel: low | maxOutputTokens: 1,024 |
| Grok 4.6 | xAI | Responses | grok-4.6 | reasoning.effort: low | max_output_tokens: 1,024 |
| DeepSeek V4 Pro | DeepSeek | Chat Completions | deepseek-v4-pro | thinking.type: disabled | max_tokens: 1,024 |
| Mistral Medium 3.5 | Mistral | Chat Completions | mistral-medium-3-5 | reasoning_effort: none | max_tokens: 1,024 |

**Panel B. Limit and usage interpretation**

| Configuration | Thought in limit / usage | Provider-native usage / limit note |
|---|---|---|
| GPT-5.6 Sol | Yes | Completion usage includes reasoning detail |
| GPT-5.6 Terra | Yes | Completion usage includes reasoning detail |
| Claude Sonnet 5 | Yes if thinking enabled | Input/output from provider usage |
| Gemini 3.8 Flash | Candidate + thought | Candidate and thought count once; low is lowest supported |
| Grok 4.6 | Not equivalent to usage bound | Output usage includes nested reasoning; low is lowest documented |
| DeepSeek V4 Pro | Not established | Cache hit/miss recorded; finish reason governs truncation |
| Mistral Medium 3.5 | Not established | None is minimal, not verified zero reasoning |

**Note.** The 1,024-token limit followed provider-native conventions; DeepSeek and Mistral cap semantics could not be verified.

**Table S4. Actual U1 single-dimension unit transformations**

| Dimension | U0 unit | U1 unit | Exact conversion | U1 items | Output request changed |
|---|---|---|---|---|---|
| SOC concentration | g C $kg^{-1}$ | % | 1% = 10 g C $kg^{-1}$ | 16 (8 bases × F0/F1) | No |
| Whole-soil bulk density | g $cm^{-3}$ | kg $m^{-3}$ | 1 g $cm^{-3}$ = 1,000 kg $m^{-3}$ | 16 (8 bases × F0/F1) | No |
| Layer thickness | cm | m | 1 m = 100 cm | 14 (7 bases × F0/F1) | No |
| Requested SOC stock output | t C $ha^{-1}$ | kg C $m^{-2}$ | 1 t C $ha^{-1}$ = 0.1 kg C $m^{-2}$ | 14 (7 bases × F0/F1) | Yes |

**Note.** Each U1 transformation affected one dimension per base scenario under both F0 and F1.

**Table S5. Response completeness, parsing and validity**

**Panel A**

| Task | Model | Condition | Planned | Returned | Valid parsed | Invalid output | Terminal failure | Strict JSON | Fallback | Invalid / terminal (%) |
|---|---|---|---|---|---|---|---|---|---|---|
| T1 | GPT-5.6 Sol | T1a | 6,160 | 6,158 | 6,158 | 0 | 2 | 6,158 | 0 | 0.032 |
| T1 | GPT-5.6 Sol | T1b | 6,160 | 6,156 | 6,156 | 0 | 4 | 6,156 | 0 | 0.065 |
| T1 | GPT-5.6 Terra | T1a | 6,160 | 6,160 | 6,160 | 0 | 0 | 6,160 | 0 | 0.000 |
| T1 | GPT-5.6 Terra | T1b | 6,160 | 6,160 | 6,160 | 0 | 0 | 6,160 | 0 | 0.000 |
| T1 | Claude Sonnet 5 | T1a | 6,160 | 6,160 | 6,160 | 0 | 0 | 6,160 | 0 | 0.000 |
| T1 | Claude Sonnet 5 | T1b | 6,160 | 6,160 | 6,160 | 0 | 0 | 6,160 | 0 | 0.000 |
| T1 | Gemini 3.8 Flash | T1a | 6,160 | 6,160 | 6,160 | 0 | 0 | 1,127 | 5,033 | 0.000 |
| T1 | Gemini 3.8 Flash | T1b | 6,160 | 6,160 | 6,160 | 0 | 0 | 2,833 | 3,327 | 0.000 |
| T1 | Grok 4.6 | T1a | 6,160 | 6,159 | 6,159 | 0 | 1 | 6,159 | 0 | 0.016 |
| T1 | Grok 4.6 | T1b | 6,160 | 6,155 | 6,155 | 0 | 5 | 6,155 | 0 | 0.081 |

| Task | Model | Condition | Planned | Returned | Valid parsed | Invalid output | Terminal failure | Strict JSON | Fallback | Invalid / terminal (%) |
|---|---|---|---|---|---|---|---|---|---|---|
| T1 | DeepSeek V4 Pro | T1a | 6,160 | 6,156 | 6,156 | 0 | 4 | 6,133 | 23 | 0.065 |
| T1 | DeepSeek V4 Pro | T1b | 6,160 | 6,160 | 6,160 | 0 | 0 | 6,151 | 9 | 0.000 |
| T1 | Mistral Medium 3.5 | T1a | 6,160 | 6,160 | 6,160 | 0 | 0 | 6,160 | 0 | 0.000 |
| T1 | Mistral Medium 3.5 | T1b | 6,160 | 6,159 | 6,159 | 0 | 1 | 6,159 | 0 | 0.016 |
| T1 | Task total | All | 86,240 | 86,223 | 86,223 | 0 | 17 | 77,831 | 8,392 | 0.020 |

**Panel B**

| Task | Model | Condition | Planned | Returned | Valid parsed | Invalid output | Terminal failure | Strict JSON | Fallback | Invalid / terminal (%) |
|---|---|---|---|---|---|---|---|---|---|---|
| T2 | GPT-5.6 Sol | U0F0 | 600 | 600 | 600 | 0 | 0 | 600 | 0 | 0.000 |
| T2 | GPT-5.6 Sol | U0F1 | 600 | 600 | 600 | 0 | 0 | 600 | 0 | 0.000 |
| T2 | GPT-5.6 Sol | U1F0 | 600 | 600 | 600 | 0 | 0 | 600 | 0 | 0.000 |
| T2 | GPT-5.6 Sol | U1F1 | 600 | 600 | 600 | 0 | 0 | 600 | 0 | 0.000 |
| T2 | GPT-5.6 Terra | U0F0 | 600 | 600 | 600 | 0 | 0 | 600 | 0 | 0.000 |
| T2 | GPT-5.6 Terra | U0F1 | 600 | 600 | 600 | 0 | 0 | 600 | 0 | 0.000 |
| T2 | GPT-5.6 Terra | U1F0 | 600 | 600 | 600 | 0 | 0 | 600 | 0 | 0.000 |
| T2 | GPT-5.6 Terra | U1F1 | 600 | 600 | 600 | 0 | 0 | 600 | 0 | 0.000 |
| T2 | Claude Sonnet 5 | U0F0 | 600 | 600 | 600 | 0 | 0 | 166 | 434 | 0.000 |
| T2 | Claude Sonnet 5 | U0F1 | 600 | 600 | 600 | 0 | 0 | 1 | 599 | 0.000 |
| T2 | Claude Sonnet 5 | U1F0 | 600 | 600 | 600 | 0 | 0 | 129 | 471 | 0.000 |
| T2 | Claude Sonnet 5 | U1F1 | 600 | 600 | 600 | 0 | 0 | 22 | 578 | 0.000 |
| T2 | Gemini 3.8 Flash | U0F0 | 600 | 600 | 600 | 0 | 0 | 51 | 549 | 0.000 |
| T2 | Gemini 3.8 Flash | U0F1 | 600 | 600 | 600 | 0 | 0 | 92 | 508 | 0.000 |
| T2 | Gemini 3.8 Flash | U1F0 | 600 | 600 | 600 | 0 | 0 | 32 | 568 | 0.000 |
| T2 | Gemini 3.8 Flash | U1F1 | 600 | 600 | 600 | 0 | 0 | 34 | 566 | 0.000 |
| T2 | Grok 4.6 | U0F0 | 600 | 600 | 600 | 0 | 0 | 600 | 0 | 0.000 |
| T2 | Grok 4.6 | U0F1 | 600 | 599 | 599 | 0 | 1 | 599 | 0 | 0.167 |
| T2 | Grok 4.6 | U1F0 | 600 | 600 | 600 | 0 | 0 | 600 | 0 | 0.000 |
| T2 | Grok 4.6 | U1F1 | 600 | 600 | 600 | 0 | 0 | 600 | 0 | 0.000 |
| T2 | DeepSeek V4 Pro | U0F0 | 600 | 600 | 600 | 0 | 0 | 600 | 0 | 0.000 |
| T2 | DeepSeek V4 Pro | U0F1 | 600 | 600 | 600 | 0 | 0 | 600 | 0 | 0.000 |
| T2 | DeepSeek V4 Pro | U1F0 | 600 | 599 | 599 | 0 | 1 | 599 | 0 | 0.167 |
| T2 | DeepSeek V4 Pro | U1F1 | 600 | 599 | 594 | 5 | 1 | 594 | 0 | 1.000 |
| T2 | Mistral Medium 3.5 | U0F0 | 600 | 600 | 600 | 0 | 0 | 600 | 0 | 0.000 |
| T2 | Mistral Medium 3.5 | U0F1 | 600 | 600 | 600 | 0 | 0 | 600 | 0 | 0.000 |

| Task | Model | Condition | Planned | Returned | Valid parsed | Invalid output | Terminal failure | Strict JSON | Fallback | Invalid / terminal (%) |
|---|---|---|---|---|---|---|---|---|---|---|
| T2 | Mistral Medium 3.5 | U1F0 | 600 | 600 | 600 | 0 | 0 | 600 | 0 | 0.000 |
| T2 | Mistral Medium 3.5 | U1F1 | 600 | 600 | 600 | 0 | 0 | 600 | 0 | 0.000 |
| T2 | Task total | All | 16,800 | 16,797 | 16,792 | 5 | 3 | 12,519 | 4,273 | 0.048 |

**Panel C**

| Task | Model | Condition | Planned | Returned | Valid parsed | Invalid output | Terminal failure | Strict JSON | Fallback | Invalid / terminal (%) |
|---|---|---|---|---|---|---|---|---|---|---|
| T3 | GPT-5.6 Sol | forward | 3,000 | 3,000 | 3,000 | 0 | 0 | 3,000 | 0 | 0.000 |
| T3 | GPT-5.6 Sol | reverse | 3,000 | 2,998 | 2,998 | 0 | 2 | 2,998 | 0 | 0.067 |
| T3 | GPT-5.6 Terra | forward | 3,000 | 3,000 | 3,000 | 0 | 0 | 3,000 | 0 | 0.000 |
| T3 | GPT-5.6 Terra | reverse | 3,000 | 3,000 | 3,000 | 0 | 0 | 3,000 | 0 | 0.000 |
| T3 | Claude Sonnet 5 | forward | 3,000 | 3,000 | 3,000 | 0 | 0 | 3,000 | 0 | 0.000 |
| T3 | Claude Sonnet 5 | reverse | 3,000 | 3,000 | 3,000 | 0 | 0 | 3,000 | 0 | 0.000 |
| T3 | Gemini 3.8 Flash | forward | 3,000 | 3,000 | 3,000 | 0 | 0 | 1,995 | 1,005 | 0.000 |
| T3 | Gemini 3.8 Flash | reverse | 3,000 | 3,000 | 3,000 | 0 | 0 | 2,037 | 963 | 0.000 |
| T3 | Grok 4.6 | forward | 3,000 | 3,000 | 3,000 | 0 | 0 | 3,000 | 0 | 0.000 |
| T3 | Grok 4.6 | reverse | 3,000 | 2,998 | 2,998 | 0 | 2 | 2,998 | 0 | 0.067 |
| T3 | DeepSeek V4 Pro | forward | 3,000 | 3,000 | 3,000 | 0 | 0 | 2,991 | 9 | 0.000 |
| T3 | DeepSeek V4 Pro | reverse | 3,000 | 3,000 | 3,000 | 0 | 0 | 2,993 | 7 | 0.000 |
| T3 | Mistral Medium 3.5 | forward | 3,000 | 3,000 | 3,000 | 0 | 0 | 2,972 | 28 | 0.000 |
| T3 | Mistral Medium 3.5 | reverse | 3,000 | 3,000 | 3,000 | 0 | 0 | 2,978 | 22 | 0.000 |
| T3 | Task total | All | 42,000 | 41,996 | 41,996 | 0 | 4 | 39,962 | 2,034 | 0.010 |

**Panel D**

| Task | Model | Condition | Planned | Returned | Valid parsed | Invalid output | Terminal failure | Strict JSON | Fallback | Invalid / terminal (%) |
|---|---|---|---|---|---|---|---|---|---|---|
| All | Grand total | All | 145,040 | 145,016 | 145,011 | 5 | 24 | 130,312 | 14,699 | 0.020 |

**Note.** Percentages use planned responses. Returned = valid + invalid; strict JSON + fallback = valid. Forward/reverse denotes canonical ordering.

**Table S6. T1 supervised baselines under spatial cross-validation**

| Reference | Predictor | MAE (g C kg$^{-1}$) | Paddocks |
|---|---|---|---|
| B0 | Training-fold SOC median | 13.4772 | 308 |
| B1 | Training-fold same-state median (B0 fallback) | 18.5128 | 308 |
| B2 | Random forest, T1a fields | 3.1338 | 308 |

| Reference | Predictor | MAE (g C kg⁻¹) | Paddocks |
|---|---|---|---|
| B2′ | Random forest, T1b fields | 14.2926 | 308 |

**Note.** All baselines used measured training labels and the same five spatial folds; results are contextual, not like-for-like closed-book comparisons.

**Table S7. T2 accuracy and matched contrasts**

**Panel A**

| Configuration | Overall (%) | U0F0 | U1F0 | U0F1 | U1F1 |
|---|---|---|---|---|---|
| GPT-5.6 Sol | 84.00 | 99.50 | 69.67 | 96.33 | 70.50 |
| GPT-5.6 Terra | 78.21 | 98.50 | 68.17 | 79.50 | 66.67 |
| Claude Sonnet 5 | 76.67 | 99.17 | 59.50 | 99.33 | 48.67 |
| Gemini 3.8 Flash | 89.00 | 100.00 | 76.67 | 100.00 | 79.33 |
| Grok 4.6 | 83.21 | 100.00 | 65.33 | 99.83 | 67.67 |
| DeepSeek V4 Pro | 76.08 | 92.83 | 61.33 | 87.50 | 62.67 |
| Mistral Medium 3.5 | 38.88 | 58.33 | 58.67 | 21.00 | 17.50 |

**Panel B**

| Configuration | Contrast | Estimate | 95% lower | 95% upper |
|---|---|---|---|---|
| GPT-5.6 Sol | U1 − U0 | -27.83 | -29.37 | -26.29 |
| GPT-5.6 Sol | F1 − F0 | -1.17 | -2.71 | 0.37 |
| GPT-5.6 Sol | U×F | 4.00 | 0.92 | 6.91 |
| GPT-5.6 Terra | U1 − U0 | -21.58 | -23.46 | -19.79 |
| GPT-5.6 Terra | F1 − F0 | -10.25 | -12.05 | -8.45 |
| GPT-5.6 Terra | U×F | 17.50 | 13.74 | 21.10 |
| Claude Sonnet 5 | U1 − U0 | -45.17 | -47.22 | -43.03 |
| Claude Sonnet 5 | F1 − F0 | -5.33 | -7.39 | -3.28 |
| Claude Sonnet 5 | U×F | -11.00 | -15.10 | -6.90 |
| Gemini 3.8 Flash | U1 − U0 | -22.00 | -23.28 | -20.63 |
| Gemini 3.8 Flash | F1 − F0 | 1.33 | 0.05 | 2.62 |
| Gemini 3.8 Flash | U×F | 2.67 | 0.10 | 5.23 |
| Grok 4.6 | U1 − U0 | -33.42 | -34.70 | -32.13 |
| Grok 4.6 | F1 − F0 | 1.08 | -0.20 | 2.37 |
| Grok 4.6 | U×F | 2.50 | -0.06 | 5.06 |
| DeepSeek V4 Pro | U1 − U0 | -28.17 | -30.65 | -25.60 |
| DeepSeek V4 Pro | F1 − F0 | -2.00 | -4.56 | 0.56 |
| DeepSeek V4 Pro | U×F | 6.67 | 1.54 | 11.80 |
| Mistral Medium 3.5 | U1 − U0 | -1.58 | -3.12 | -0.04 |

| Configuration | Contrast | Estimate | 95% lower | 95% upper |
|---|---|---|---|---|
| Mistral Medium 3.5 | F1 − F0 | -39.25 | -40.79 | -37.71 |
| Mistral Medium 3.5 | U×F | -3.83 | -6.91 | -0.76 |

**Note.** Overall accuracy equally weights 120 items. Contrasts average across 30 bases and the other factor. Intervals reflect repeated calls on fixed items.

**Table S8. T3 fitted position effects and direct position accuracy**

**Panel A**

| Configuration | Estimate | 95% lower | 95% upper |
|---|---|---|---|
| GPT-5.6 Sol | 1.81 | -0.21 | 3.84 |
| GPT-5.6 Terra | -1.63 | -3.55 | 0.29 |
| Claude Sonnet 5 | 1.51 | -0.50 | 3.51 |
| Gemini 3.8 Flash | 2.67 | 0.11 | 5.22 |
| Grok 4.6 | 3.40 | -0.02 | 6.82 |
| DeepSeek V4 Pro | -0.34 | -1.07 | 0.39 |
| Mistral Medium 3.5 | 18.10 | 9.77 | 26.43 |

**Panel B**

| Configuration | Higher-SOC at A | Higher-SOC at B | Planned A | Planned B |
|---|---|---|---|---|
| GPT-5.6 Sol | 94.57 | 93.50 | 3,000 | 3,000 |
| GPT-5.6 Terra | 93.67 | 94.63 | 3,000 | 3,000 |
| Claude Sonnet 5 | 94.67 | 93.93 | 3,000 | 3,000 |
| Gemini 3.8 Flash | 95.10 | 94.07 | 3,000 | 3,000 |
| Grok 4.6 | 95.13 | 92.17 | 3,000 | 3,000 |
| DeepSeek V4 Pro | 93.30 | 93.63 | 3,000 | 3,000 |
| Mistral Medium 3.5 | 98.70 | 82.23 | 3,000 | 3,000 |

**Note.** Panel A reports model-based conditional A − B effects with random effects set to zero; Panel B reports direct position-specific accuracies.

**Table S9. Prespecified T3 pairwise direct-accuracy differences**

| Configuration 1 | Configuration 2 | Difference (pp) | 95% lower (pp) | 95% upper (pp) | Holm-adjusted p |
|---|---|---|---|---|---|
| GPT-5.6 Sol | GPT-5.6 Terra | -0.12 | -0.35 | 0.13 | 0.6766 |
| GPT-5.6 Sol | Claude Sonnet 5 | -0.27 | -0.44 | -0.07 | 0.0193 |
| GPT-5.6 Sol | Gemini 3.8 Flash | -0.55 | -0.76 | -0.34 | <0.0001 |
| GPT-5.6 Sol | Grok 4.6 | 0.38 | 0.08 | 0.68 | 0.0450 |
| GPT-5.6 Sol | DeepSeek V4 Pro | 0.57 | 0.23 | 0.90 | 0.0069 |
| GPT-5.6 Sol | Mistral Medium 3.5 | 3.57 | 3.34 | 3.80 | <0.0001 |
| GPT-5.6 Terra | Claude Sonnet 5 | -0.15 | -0.33 | 0.03 | 0.2916 |

| Configuration 1 | Configuration 2 | Difference (pp) | 95% lower (pp) | 95% upper (pp) | Holm-adjusted p |
|---|---|---|---|---|---|
| GPT-5.6 Terra | Gemini 3.8 Flash | -0.43 | -0.63 | -0.24 | 0.0002 |
| GPT-5.6 Terra | Grok 4.6 | 0.50 | 0.22 | 0.80 | 0.0061 |
| GPT-5.6 Terra | DeepSeek V4 Pro | 0.68 | 0.35 | 1.02 | 0.0006 |
| GPT-5.6 Terra | Mistral Medium 3.5 | 3.68 | 3.45 | 3.91 | <0.0001 |
| Claude Sonnet 5 | Gemini 3.8 Flash | -0.28 | -0.41 | -0.16 | 0.0001 |
| Claude Sonnet 5 | Grok 4.6 | 0.65 | 0.40 | 0.90 | <0.0001 |
| Claude Sonnet 5 | DeepSeek V4 Pro | 0.83 | 0.53 | 1.13 | <0.0001 |
| Claude Sonnet 5 | Mistral Medium 3.5 | 3.83 | 3.68 | 3.99 | <0.0001 |
| Gemini 3.8 Flash | Grok 4.6 | 0.93 | 0.67 | 1.20 | <0.0001 |
| Gemini 3.8 Flash | DeepSeek V4 Pro | 1.12 | 0.80 | 1.43 | <0.0001 |
| Gemini 3.8 Flash | Mistral Medium 3.5 | 4.12 | 3.92 | 4.31 | <0.0001 |
| Grok 4.6 | DeepSeek V4 Pro | 0.18 | -0.20 | 0.55 | 0.6766 |
| Grok 4.6 | Mistral Medium 3.5 | 3.18 | 2.90 | 3.46 | <0.0001 |
| DeepSeek V4 Pro | Mistral Medium 3.5 | 3.00 | 2.67 | 3.33 | <0.0001 |

**Note.** Differences are configuration 1 minus configuration 2 (percentage points). Holm adjustment applies to the 21 P values, not interval endpoints.

**Table S10. T3 nominal-separation sensitivity**

| Configuration | All 300 pairs (%) | 248 meeting threshold (%) | 52 complement (%) |
|---|---|---|---|
| GPT-5.6 Sol | 94.03 | 99.31 | 68.85 |
| GPT-5.6 Terra | 94.15 | 99.27 | 69.71 |
| Claude Sonnet 5 | 94.30 | 99.54 | 69.33 |
| Gemini 3.8 Flash | 94.58 | 99.56 | 70.87 |
| Grok 4.6 | 93.65 | 98.83 | 68.94 |
| DeepSeek V4 Pro | 93.47 | 99.17 | 66.25 |
| Mistral Medium 3.5 | 90.47 | 97.04 | 59.13 |

**Note.** The 248/52 split was applied after random selection; the 10% rule is not a measured uncertainty estimate.

**Table S11. Post-collection T1 and T2 numerical-answer IQR distributions**

**Panel A**

| Configuration | Condition | Cells | Zero | Mean | P5 | Q1 | P50 | Q3 | P95 | Max |
|---|---|---|---|---|---|---|---|---|---|---|
| GPT-5.6 Sol | T1a | 308 | 3 | 1.9318 | 0.2000 | 0.7437 | 1.2000 | 2.5625 | 5.3150 | 16.6000 |
| GPT-5.6 Sol | T1b | 308 | 15 | 3.0120 | 0.0250 | 0.9000 | 1.9000 | 3.2000 | 9.7450 | 31.4250 |
| GPT-5.6 Terra | T1a | 308 | 42 | 1.5793 | 0.0000 | 0.3000 | 1.0000 | 2.1750 | 4.4812 | 16.5250 |
| GPT-5.6 Terra | T1b | 308 | 12 | 2.5119 | 0.0750 | 0.4000 | 2.0000 | 3.7000 | 5.8212 | 28.1500 |

| Configuration | Condition | Cells | Zero | Mean | P5 | Q1 | P50 | Q3 | P95 | Max |
|---|---|---|---|---|---|---|---|---|---|---|
| Claude Sonnet 5 | T1a | 308 | 173 | 0.5093 | 0.0000 | 0.0000 | 0.0000 | 0.6000 | 2.6300 | 15.6250 |
| Claude Sonnet 5 | T1b | 308 | 200 | 0.6562 | 0.0000 | 0.0000 | 0.0000 | 0.7000 | 3.7550 | 10.0000 |
| Gemini 3.8 Flash | T1a | 308 | 21 | 0.5615 | 0.0000 | 0.1937 | 0.3200 | 0.7750 | 1.7000 | 6.9300 |
| Gemini 3.8 Flash | T1b | 308 | 11 | 2.3593 | 0.2587 | 1.0000 | 1.7000 | 3.2500 | 6.2762 | 17.0250 |
| Grok 4.6 | T1a | 308 | 10 | 1.2823 | 0.0250 | 0.2750 | 0.7000 | 1.6000 | 4.6488 | 8.8500 |
| Grok 4.6 | T1b | 308 | 0 | 2.5991 | 0.5000 | 1.2000 | 2.0500 | 3.6250 | 6.2000 | 10.7500 |
| DeepSeek V4 Pro | T1a | 308 | 0 | 6.8263 | 1.1951 | 2.3231 | 4.4500 | 8.6331 | 21.9325 | 47.1900 |
| DeepSeek V4 Pro | T1b | 308 | 0 | 4.5053 | 1.0544 | 2.0000 | 3.3500 | 6.4813 | 11.3587 | 17.1500 |
| Mistral Medium 3.5 | T1a | 308 | 171 | 0.8295 | 0.0000 | 0.0000 | 0.0000 | 0.5000 | 4.0215 | 11.6600 |
| Mistral Medium 3.5 | T1b | 308 | 237 | 0.8423 | 0.0000 | 0.0000 | 0.0000 | 0.0000 | 6.8000 | 13.0000 |

**Panel B**

| Configuration | Condition | Cells | Zero | Mean | P5 | Q1 | P50 | Q3 | P95 | Max |
|---|---|---|---|---|---|---|---|---|---|---|
| GPT-5.6 Sol | all items | 120 | 29 | 56.9514 | 0.0000 | 0.0005 | 0.0152 | 0.0809 | 898.9149 | 900.0055 |
| GPT-5.6 Terra | all items | 120 | 14 | 10.5847 | 0.0000 | 0.0064 | 0.0276 | 0.2576 | 89.9698 | 225.2513 |
| Claude Sonnet 5 | all items | 120 | 29 | 6783.6135 | 0.0000 | 0.0009 | 0.0254 | 9.0028 | 98970.3358 | 99900.6057 |
| Gemini 3.8 Flash | all items | 120 | 46 | 6.3797 | 0.0000 | 0.0000 | 0.0008 | 0.0076 | 89.9940 | 90.0031 |
| Grok 4.6 | all items | 120 | 75 | 6.1881 | 0.0000 | 0.0000 | 0.0000 | 0.0005 | 98.9893 | 99.0000 |
| DeepSeek V4 Pro | all items | 120 | 13 | 59.4112 | 0.0000 | 0.0286 | 0.1234 | 0.5140 | 263.9082 | 900.2087 |
| Mistral Medium 3.5 | all items | 120 | 10 | 23.0274 | 0.0000 | 0.0326 | 0.4432 | 2.3395 | 126.0776 | 894.0993 |

**Note.** Cells have equal weight. T1 IQR is in g C $kg^{-1}$; T2 IQR is a percentage of the absolute reference stock. Quartiles use linear interpolation; zeros and valid extremes are retained.

**Table S12. T3 within-order repeated-choice disagreement**

| Configuration | Order | Planned | Estimable | Zero | Nonzero | Nonzero (%) | Unestimable |
|---|---|---|---|---|---|---|---|
| GPT-5.6 Sol | forward | 300 | 300 | 294 | 6 | 2.0 | 0 |
| GPT-5.6 Sol | reverse | 300 | 300 | 293 | 7 | 2.333333 | 0 |
| GPT-5.6 Terra | forward | 300 | 300 | 293 | 7 | 2.333333 | 0 |
| GPT-5.6 Terra | reverse | 300 | 300 | 294 | 6 | 2.0 | 0 |
| Claude Sonnet 5 | forward | 300 | 300 | 299 | 1 | 0.333333 | 0 |
| Claude Sonnet 5 | reverse | 300 | 300 | 299 | 1 | 0.333333 | 0 |
| Gemini 3.8 Flash | forward | 300 | 300 | 299 | 1 | 0.333333 | 0 |
| Gemini 3.8 Flash | reverse | 300 | 300 | 295 | 5 | 1.666667 | 0 |
| Grok 4.6 | forward | 300 | 300 | 283 | 17 | 5.666667 | 0 |
| Grok 4.6 | reverse | 300 | 300 | 287 | 13 | 4.333333 | 0 |
| DeepSeek V4 Pro | forward | 300 | 300 | 274 | 26 | 8.666667 | 0 |
| DeepSeek V4 Pro | reverse | 300 | 300 | 268 | 32 | 10.666667 | 0 |
| Mistral Medium 3.5 | forward | 300 | 300 | 294 | 6 | 2.0 | 0 |
| Mistral Medium 3.5 | reverse | 300 | 300 | 294 | 6 | 2.0 | 0 |

**Note.** Forward/reverse denotes canonical order. Nonzero percentages use estimable cells as the denominator; zero disagreement does not imply correctness.

**Table S13. Provider-reported token use and estimated API cost**

**Panel A**

| Configuration | Task | Input | Provider output | Billable output | Cache hit | Cache miss | Reasoning | Thought |
|---|---|---|---|---|---|---|---|---|
| GPT-5.6 Sol | T1 | 2,121,672 | 125,852 | 125,852 | 0 | 0 | 0 | 0 |
| GPT-5.6 Sol | T2 | 448,960 | 26,321 | 26,321 | 0 | 0 | 0 | 0 |
| GPT-5.6 Sol | T3 | 2,478,135 | 47,984 | 47,984 | 0 | 0 | 0 | 0 |
| GPT-5.6 Terra | T1 | 2,122,600 | 123,562 | 123,562 | 0 | 0 | 0 | 0 |
| GPT-5.6 Terra | T2 | 448,960 | 30,302 | 30,302 | 0 | 0 | 0 | 0 |
| GPT-5.6 Terra | T3 | 2,478,960 | 48,000 | 48,000 | 0 | 0 | 0 | 0 |
| Claude Sonnet 5 | T1 | 3,251,280 | 158,790 | 158,790 | 0 | 0 | 0 | 0 |
| Claude Sonnet 5 | T2 | 717,720 | 413,285 | 413,285 | 0 | 0 | 0 | 0 |
| Claude Sonnet 5 | T3 | 3,818,100 | 72,000 | 72,000 | 0 | 0 | 0 | 0 |
| Gemini 3.8 Flash | T1 | 2,397,900 | 150,102 | 1,539,785 | 0 | 0 | 0 | 1,389,683 |
| Gemini 3.8 Flash | T2 | 459,360 | 38,073 | 1,044,838 | 0 | 0 | 0 | 1,006,765 |
| Gemini 3.8 Flash | T3 | 2,956,100 | 45,359 | 739,846 | 0 | 0 | 0 | 694,487 |
| Grok 4.6 | T1 | 9,904,283 | 4,325,633 | 4,325,633 | 0 | 0 | 4,231,493 | 0 |

| Configuration | Task | Input | Provider output | Billable output | Cache hit | Cache miss | Reasoning | Thought |
|---|---|---|---|---|---|---|---|---|
| Grok 4.6 | T2 | 1,960,703 | 1,809,980 | 1,809,980 | 0 | 0 | 1,788,590 | 0 |
| Grok 4.6 | T3 | 6,286,902 | 1,299,836 | 1,299,836 | 0 | 0 | 1,263,848 | 0 |
| DeepSeek V4 Pro | T1 | 2,100,605 | 111,468 | 111,468 | 748,672 | 1,351,933 | 0 | 0 |
| DeepSeek V4 Pro | T2 | 427,564 | 27,458 | 27,458 | 301,312 | 126,252 | 0 | 0 |
| DeepSeek V4 Pro | T3 | 2,458,180 | 42,591 | 42,591 | 2,074,368 | 383,812 | 0 | 0 |
| Mistral Medium 3.5 | T1 | 2,610,682 | 135,632 | 135,632 | 768 | 0 | 0 | 0 |
| Mistral Medium 3.5 | T2 | 490,000 | 32,124 | 32,124 | 19,584 | 0 | 0 | 0 |
| Mistral Medium 3.5 | T3 | 3,061,240 | 48,200 | 48,200 | 1,024 | 0 | 0 | 0 |

**Panel B**

| Configuration | T1 | T2 | T3 | Total | Reservation | Retry cost |
|---|---|---|---|---|---|---|
| GPT-5.6 Sol | 11.00 | 2.32 | 10.87 | 24.20 | 0.92 | 0.00 |
| GPT-5.6 Terra | 5.73 | 1.26 | 5.53 | 12.52 | 0.00 | 0.00 |
| Claude Sonnet 5 | 8.09 | 5.57 | 8.36 | 22.01 | 0.03 | 0.00 |
| Gemini 3.8 Flash | 7.57 | 4.26 | 4.99 | 16.83 | 0.09 | 0.01 |
| Grok 4.6 | 45.76 | 14.78 | 20.37 | 80.92 | 0.31 | 0.05 |
| DeepSeek V4 Pro | 3.21 | 0.67 | 3.41 | 7.30 | 0.17 | 0.00 |
| Mistral Medium 3.5 | 4.93 | 0.98 | 4.95 | 10.86 | 0.04 | 0.00 |
| All configurations | 86.30 | 29.85 | 58.49 | 174.64 | 1.55 | 0.05 |

**Note.** Reasoning/thought counts may already be included in output totals and should not be added twice. Confirmed cost includes priced retries; unknown usage remains a separate reservation.

**Table S14. Frozen 29 September 2026 provider price register**

| Configuration | Provider | Input US$/1M | Output US$/1M | Cached-input rate | Checked | Archived source URL |
|---|---|---|---|---|---|---|
| GPT-5.6 Sol | OpenAI | 4 | 20 | Not separately applied | 2026-09-29 | https://developers.openai.com/api/docs/models/gpt-5.6-sol |
| GPT-5.6 Terra | OpenAI | 2 | 12 | Not separately applied | 2026-09-29 | https://developers.openai.com/api/docs/models |
| Claude Sonnet 5 | Anthropic | 2 | 10 | Not separately applied | 2026-09-29 | https://www.anthropic.com/news/claude-sonnet-5 |
| Gemini 3.8 Flash | Google | 0.75 | 3.75 | Not separately applied | 2026-09-29 | https://ai.google.dev/gemini-api/docs/latest-model |
| Grok 4.6 | xAI | 2 | 6 | Not separately applied | 2026-09-29 | https://docs.x.ai/developers/models/grok-4.6 |
| DeepSeek V4 Pro | DeepSeek | 1.32 | 3.96 | Not separately applied | 2026-09-29 | https://api-docs.deepseek.com/quick_start/pricing/ |
| Mistral Medium 3.5 | Mistral | 1.5 | 7.5 | Not separately applied | 2026-09-29 | https://docs.mistral.ai/models/mistral-medium-3-5-26-04 |

**Note.** Rates are US$ per million tokens (29 September 2026). The estimator did not apply separate cached-input discounts; unknown rate details are not inferred.

### Supplementary Figure

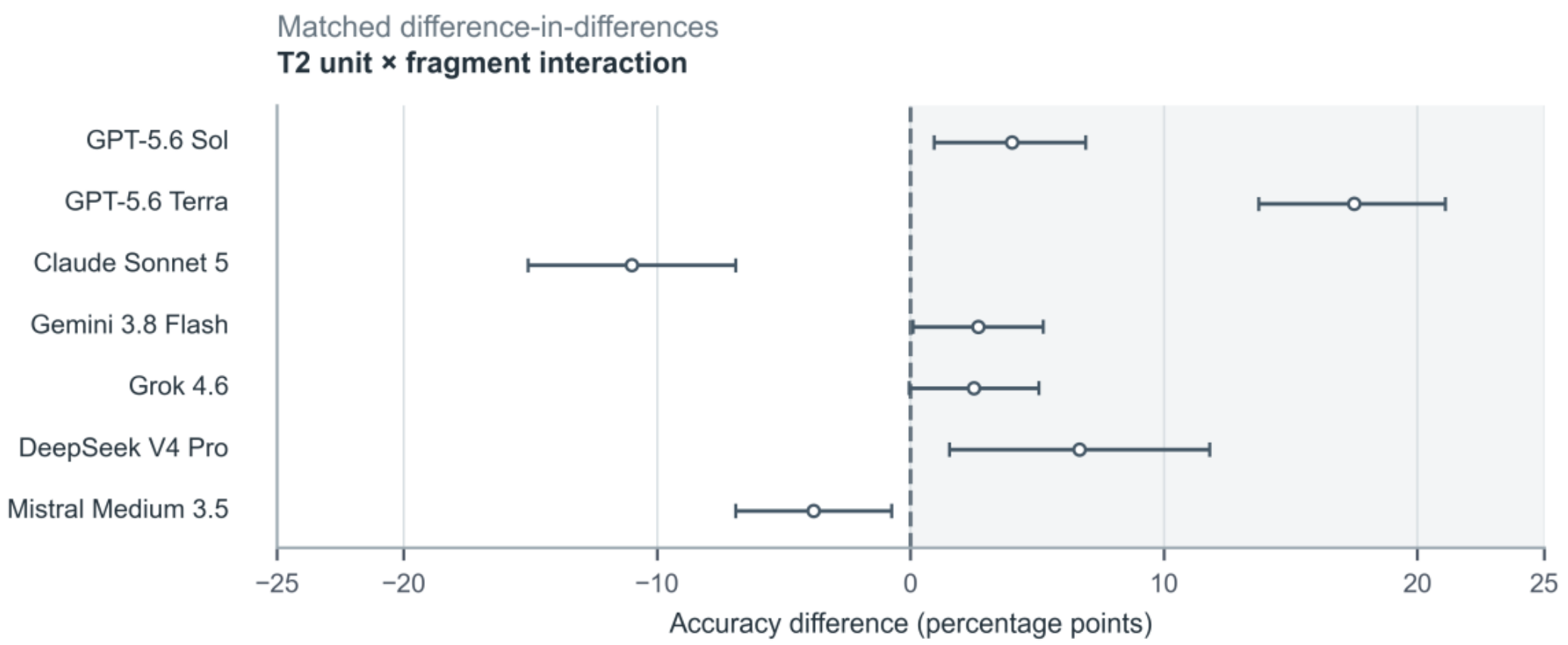


**Figure S1. Unit × coarse-fragment interaction in T2.** Differences-in-differences are (U1F1 − U0F1) − (U1F0 − U0F0), averaged across 30 base scenarios. Symbols show estimated interactions and pointwise 95% repeated-call intervals (percentage points); the dashed line marks zero. A positive interaction means the U1 − U0 accuracy contrast is more positive under F1 than F0, not necessarily that accuracy is higher. The theoretical range is −200 to +200 percentage points; the clipping limitation is noted in M6.

### Supplementary References

### Data and code availability

SOC-M v1 is publicly available (DOI: 10.25919/z378-4286). The 120 T2 scenarios and displayed-value references are provided as Supplementary Data with this document. Additional benchmark items, prompts, analysis code and collection records remain in the Project 2 archive; no public repository or code DOI has yet been established for these materials. FarmData coordinates, authentication credentials and raw response ledgers are not included. Any further release requires author approval and privacy review.


\end{document}